\documentclass[iop,usenatbib]{emulateapj}

\usepackage[english]{babel}
\usepackage{graphicx}
\usepackage{fleqn}
\usepackage{amssymb}
\usepackage{amsmath}
\usepackage{booktabs}
\usepackage{comment}

\usepackage[breaklinks,colorlinks,
   urlcolor=blue,citecolor=blue,linkcolor=blue]{hyperref}
\usepackage{comment}

\usepackage{color, colortbl}

\usepackage{txfonts}

\newcommand*{\mycdot}{\kern-.2em\cdot\kern-.2em}
\renewcommand{\S}{Section}
\newcommand{\F}{Fig.}
\newcommand{\codename}{SecularMultiple}

\newcommand{\ve}[1]{\boldsymbol{#1}}

\newcommand{\msun}{\mathrm{M}_\odot}

\newcommand{\au}{\,\textsc{au}}
\newcommand{\renc}{R_\mathrm{enc}}

\newcommand{\kms}{\mathrm{km \, s^{-1}}}

\newcommand\kick{\mathrm{k}}

\allowdisplaybreaks

\begin{document}

\title{The impact of white dwarf natal kicks and stellar flybys on the rates of Type Ia supernovae in triple-star systems}
\author{Adrian S. Hamers$^{1}$ and Todd A. Thompson$^{2,3,1}$}
\affil{$^{1}$Institute for Advanced Study, School of Natural Sciences, Einstein Drive, Princeton, NJ 08540, USA \\
$^{2}$Department of Astronomy, The Ohio State University, Columbus, Ohio 43210, USA \\
$^{3}$Center for Cosmology and AstroParticle Physics, Department of Physics, The Ohio State University, Columbus, Ohio 43210, USA}



\begin{abstract} 
Type Ia supernovae (SNe Ia) could arise from mergers of carbon-oxygen white dwarfs (WDs) triggered by Lidov-Kozai (LK) oscillations in hierarchical triple-star systems. However,  predicted merger rates are several orders of magnitude lower than the observed SNe Ia rate. The low predicted rates can be attributed in part to the fact that many potential WD-merger progenitor systems, with high mutual orbital inclination, merge or interact before the WD stage. Recently, evidence was found for the existence of natal kicks imparted on WDs with a typical magnitude of $0.75\,\kms$. In triples, kicks change the mutual inclination and in general increase the outer orbit eccentricity, bringing the triple into an active LK regime at late stages and avoiding the issue of pre-WD merger or interaction. Stars passing by the triple can result in similar effects. However, both processes can also disrupt the triple. In this paper, we quantitatively investigate the impact of WD kicks and flybys on the rate of WD mergers using detailed simulations. We find that WD kicks and flybys combine to increase the predicted WD merger rates by a factor of $\sim 2.5$, resulting in a time-integrated rate of $\approx 1.1 \times 10^{-4}\,\msun^{-1}$. Despite the significant boost, the predicted rates are still more than one order of magnitude below the observed rate of $\sim 10^{-3} \, \msun^{-1}$. However, many systematic uncertainties still remain in our calculations, in particular the potential contributions from tighter triples, dynamically unstable systems, unbound systems due to WD kicks, and quadruple systems. 
\end{abstract}


\section{Introduction}
\label{sect:introduction}
Despite their importance for distance determination on cosmological scales \citep{1998AJ....116.1009R,1999ApJ...517..565P} and decades of study, the origin of Type Ia supernovae (SNe Ia) is still not fully understood (see, e.g., \citealt{2012NewAR..56..122W,2014ARA&A..52..107M,2018arXiv180203125L} for reviews). The common view is that SNe Ia result from runaway thermonuclear explosions of degenerate carbon-oxygen (CO) white dwarfs (WDs). These explosions could be triggered in the `single degenerate' (SD) channel \citep{1973ApJ...186.1007W,1984ApJ...286..644N}, in which a CO WD accretes mass from a companion star until the WD exceeds the critical Chandrasekhar mass. Alternatively, in the `double degenerate' (DD) channel \citep{1984ApJS...54..335I,1984ApJ...277..355W}, two WDs merge due to gravitational wave emission in a tight orbit following common-envelope (CE) evolution. Both channels are confronted with challenges. Notably, there exists no evidence for the presence of mass donors in the SD channel (e.g., \citealt{2019arXiv190305115T}), and the (binary) DD channel predicts rates that are too low compared to observations (e.g., \citealt{2011MNRAS.417..408R,2014A&A...563A..83C}). 

Therefore, alternative channels for SNe Ia have been given much recent attention. A channel of particular interest is the WD collision channel, in which two WDs collide (nearly) head-on. The high temperatures and shocks during such a collision are conducive for triggering an explosion (e.g., \citealt{2009MNRAS.399L.156R,2009ApJ...705L.128R,2010ApJ...724..111R,2012ApJ...747L..10P,2015ApJ...807..105S}). The traditional scenario for colliding WDs, in dense stellar systems such as globular clusters, yields rates that are too low compared to the observed SNe Ia rate (e.g., \citealt{2009ApJ...705L.128R}). An alternative setting is triple-star systems, in which two WDs are driven to a highly eccentric orbit through Lidov-Kozai (LK) oscillations (\citealt{1962P&SS....9..719L,1962AJ.....67..591K}; see \citealt{2016ARA&A..54..441N} for a review) caused by a more distant tertiary object, resulting in a gravitational wave (GW)-driven merger \citep{2011ApJ...741...82T}, or a collision. It was argued by \citet{2012arXiv1211.4584K} that perhaps all observed SNe Ia could arise from collisions in triple systems. However, as shown by \citet{2013MNRAS.430.2262H}, this conclusion seems rather unlikely given that high mutual inclinations are needed to induce WD collisions, and the stellar progenitors in such systems would likely have merged before WD formation. This problem was later revisited by \citet{2018A&A...610A..22T}, who took into account the fact that some of the systems can enter the `semisecular' regime in which the orbit-averaging approximation breaks down \citep{2012arXiv1211.4584K,2014MNRAS.438..573B,2014ApJ...781...45A,2014MNRAS.439.1079A}. Nevertheless, \citet{2018A&A...610A..22T} came to the same conclusion that it is unlikely that all observed SNe Ia arise from WD collisions in WD triples, because the predicted rate is much lower than the observed rate.

A number of modifications to the `triple WD collision' scenario exist that could potentially increase the predicted rates. \citet{2018MNRAS.476.4234F} considered GW-driven mergers and WD collisions in quadruple-star systems composed of two binaries orbiting each other's center of mass (the `2+2' configuration). In such quadruple systems, the parameter space for exciting high eccentricities (and therefore collisions) is larger compared to equivalent triples \citep{2013MNRAS.435..943P,2017MNRAS.470.1657H,2018MNRAS.474.3547G}. However, \citet{2018MNRAS.476.4234F} did not take into account the pre-WD evolution. The pre-WD evolution was included by \citet{2018MNRAS.478..620H}, who considered both the 2+2 and 3+1 configurations (the latter consisting of a triple orbited by a fourth body, \citealt{2015MNRAS.449.4221H}), and found that both configurations yield rates that are several orders of magnitude lower compared to the observations. 

Another possibility is that the rich dynamics of triple and quadruple systems do not drive WD collisions or mergers directly, but instead are important for SNe Ia by producing more tight stellar binaries and/or WD-stellar binaries after a first episode of stellar evolution \citep{1979A&A....77..145M,2003ApJ...589..605W,2007ApJ...669.1298F,2013MNRAS.430.2262H,2013ApJ...766...64S,2014ApJ...793..137N}. These systems would then evolve as dynamically isolated tight binaries, possibly leading to WD mergers via classical binary stellar evolution (e.g., \citealt{2011MNRAS.417..408R}). Additionally, field triple systems that become dynamically unstable as a result of mass loss might lead to stellar and WD mergers \citep{2012ApJ...760...99P}. 

Yet other effects could potentially increase the production of systems that might eventually lead to SNe Ia in triples. These include gravitational perturbations from stars passing the triple system. Such perturbations can increase the mutual inclination between the inner and outer orbits, and/or the outer orbit eccentricity (e.g., \citealt{2016MNRAS.456.4219A}). Consequently, a `non-interacting' triple with inner binary components not yet evolved to WDs, could be brought into an `active' regime associated with high eccentricities and mergers at later stages when the inner binary components have evolved to WDs. Although these `flybys' effects were taken into account for quadruples by \citet{2018MNRAS.478..620H}, they have not been considered in the context of triples. 

Furthermore, \citet{2018MNRAS.480.4884E} recently constructed a catalogue of wide main sequence (MS) and WD binaries in the field using {\it Gaia} DR2 data. \citet{2018MNRAS.480.4884E} found breaks in the separation distribution of MS-WD and WD-WD binaries, which could be explained if WDs incur a natal velocity kick at their formation, with a magnitude of typically $0.75\,\kms$. Such a kick could be the result of asymmetric mass loss \citep{2003ApJ...595L..53F}. Analogously to perturbations from passing stars, natal WD kicks could potentially activate passive triples after stellar evolution, leading to higher WD merger rates. However, WD kicks and flybys can also disrupt the triple system. It is therefore not a priori clear if these processes lead to a net decrease or increase in the WD merger rates. 

In this paper, we investigate quantitatively, using detailed simulations, whether WD natal kicks and flybys could increase the WD merger rate in triples significantly enough to boost the predicted SNe Ia rate to the level of the observed rate. We describe our numerical algorithm and assumptions of the initial distributions in \S\,\ref{sect:meth}. We present our results in \S\,\ref{sect:results}, give a discussion in \S\,\ref{sect:discussion:compact}, and conclude in \S\,\ref{sect:conclusions}.

\section{Numerical approach and simulation setup}
\label{sect:meth}

\begin{table*}
\begin{tabular}{lp{6.0cm}p{8.0cm}}
\toprule
Symbol & Description & Initial value(s) and/or distribution in population synthesis \\
\midrule
{\it Triple system} \\
$m_1$					& Mass of the primary star.																	& $1-6.5\,\msun$ with a Kroupa initial mass function (\citealt{1993MNRAS.262..545K}, i.e., $\mathrm{d} N/\mathrm{d}m_1 \propto m_1^{-2.7}$ in this mass range). \\
$m_2$					& Mass of the secondary star. 																	& $m_1 q_1$, where $q_1$ has a flat distribution, and with $m_2>1\,\msun$. \\
$m_3$ 					& Mass of the tertiary star.																		& $q_2 (m_1+m_2)$, where $q_2$ has a flat distribution between 0.01 and 1, and with $m_3>0.1\,\msun$. \\
$Z_i$					& Metallicity of star $i$.																		& $0.02$ \\
$R_i$					& Radius of star $i$.																			& From stellar evolution code. \\
$P_{\mathrm{s},i}$			& Spin period of star $i$. 																		& $10\,\mathrm{d}$ \\
$\theta_{\mathrm{s},i}$		& Obliquity (spin-orbit angle) of star $i$. 															& $0^\circ$ \\
$t_{\mathrm{V},i}$			& Viscous time-scale of star $i$. 																& Computed from the stellar properties using the prescription of \citet{2002MNRAS.329..897H}. \\
$k_{\mathrm{AM},i}$			& Apsidal motion constant of star $i$.															& 0.014 \\
$r_{\mathrm{g},i}$			& Gyration radius of star $i$.																	& 0.08 \\
$\ve{V}_\kick$				& WD kick velocity.																			& Maxwellian distribution with $\sigma_\kick=0.5\,\kms$. \\
$P_{\mathrm{orb},i}$			& Orbital period of orbit $i$ (inner orbit: $i=1$; outer orbit: $i=2$).										& Gaussian distribution in $\mathrm{log}_{10}(P_{\mathrm{orb},i}/\mathrm{d})$ with mean 5.03 and standard deviation 2.28 \citep{2010ApJS..190....1R}, subject to dynamical stability \citep{2001MNRAS.321..398M}, and $a_1(1-e_1^2) > 12 \, \au$. \\
$a_i$					& Semimajor axis of orbit $i$. 																	& Computed from $P_{\mathrm{orb},i}$ and the $m_i$ using Kepler's law. \\
$e_i$					& Eccentricity of orbit $i$. 																		& Rayleigh distribution between 0.01 and 0.9 with an rms width of 0.33 \citep{2010ApJS..190....1R}. \\
$i_i$						& Inclination of orbit $i$. 																		& $0-180^\circ$ (flat distribution in $\cos i_i $) \\
$\omega_i$				& Argument of periapsis of orbit $i$. 																& $0-360^\circ$ (flat distribution in $\omega_i$) \\
$\Omega_i$				& Longitude of the ascending node of orbit $i$. 														& $0-360^\circ$ (flat distribution in $\Omega_i$) \\
\midrule
{\it Flybys} \\
$M_\mathrm{per}$			& Mass of the perturbers. 																		& $0.1-80\,\msun$ with a Kroupa initial mass function \citep{1993MNRAS.262..545K}, corrected for gravitational focusing and a stellar age of 10 Gyr. \\
$n_\star$					& Stellar number density. 																		& $0.1 \, \mathrm{pc^{-3}}$ \citep{2000MNRAS.313..209H} \\
$\renc$					& Encounter sphere radius.																	& $0, 10^4\,\au$ \\
$\sigma_\star$				& One-dimensional stellar velocity dispersion.														& $30\,\mathrm{km\,s^{-1}}$ \\
\bottomrule
\end{tabular}
\caption{Description of important quantities and their initial value(s) and/or distributions assumed in the population synthesis. }
\label{table:IC}
\end{table*}

Our numerical algorithm is implemented within the \textsc{AMUSE} framework \citep{2013CoPhC.183..456P,2013A&A...557A..84P}, and is similar to that used for quadruple-star systems by \citet{2018MNRAS.478..620H}, in this case simplifying to the case of three stars. For completeness, we give a brief summary of the most important ingredients of the algorithm, and the assumptions made. For more details, we refer to \citet{2018MNRAS.478..620H}. We give an overview of our notation in this paper in the first two columns of Table~\ref{table:IC}.

\subsection{Secular dynamical evolution}
\label{sect:meth:sec}
We describe the long-term secular dynamical evolution of the triple system using \textsc{\codename} \citep{2016MNRAS.459.2827H,2018MNRAS.476.4139H}. This code is based on an expansion of the Hamiltonian of the system in terms of ratios of separations of binaries on adjacent levels. In our case, there is only one such ratio $x\equiv r_1/r_2$, which is the ratio of the inner to the outer binary separation. The Hamiltonian is averaged over both orbits, and the equations of motion are solved numerically. We include terms up to and including fifth order in $x$ (`dotriacontupole' order). Post-Newtonian (PN) corrections are included in the inner and outer orbits to the 1PN and 2.5PN orders (i.e., relativistic precession, and energy and angular momentum loss due to gravitational wave radiation). We neglect any `cross' terms that are associated with both orbits \citep{2013ApJ...773..187N}. Note that, to the octupole order, \textsc{\codename} reduces to the technique used extensively for hierarchical triple systems in past works (e.g., \citealt{1962P&SS....9..719L,1962AJ.....67..591K,1968AJ.....73..190H,2000ApJ...535..385F,2002ApJ...578..775B,2013MNRAS.431.2155N}).

\subsection{Stellar evolution}
\label{sect:meth:stev}
To model the effects of stellar evolution, we couple the \textsc{\codename}  code within \textsc{AMUSE} with \textsc{BSE} \citep{2000MNRAS.315..543H,2002MNRAS.329..897H}, which uses analytic fits to detailed stellar evolution calculations. Output data used from \textsc{BSE} are the (convective) masses and radii of the stars at each stage in their evolution. We assume a stellar metallicity of $Z_i=0.02$ for all stars. The stars are initiated on the zero-age main sequence (MS). We neglect any pre-MS evolution (see, e.g., \citealt{2018ApJ...854...44M} for a study of the latter in triples). 

To compute the effects of stellar mass loss on the inner and outer orbits, we assume isotropic and adiabatic mass loss, i.e., $a_i M_i$ and $e_i$ are constant \citep{1956AJ.....61...49H,1963ApJ...138..471H} for both the inner and outer orbits (labeled with subscripts $i=1$ and $i=2$, respectively). Here, $M_1=m_1+m_2$ for the inner orbit, and $M_2=M_1+m_3$ for the outer orbit. 

We track the spins of the three stars, and assume that spin angular momentum is conserved as the masses and radii change due to stellar evolution (in both magnitude and direction). Note that this implies significant spin-down and spin-up when stars become giants and WDs, respectively. If conservation of spin angular momentum would imply WD spin rotation beyond critical rotation, then we set the WD spin to the critical rotation rate. Note, however, that WD spins are only relevant for WD tides, and the latter are not important in our simulations.

\subsection{Tidal evolution}
\label{sect:meth:tides}
We model the tidal evolution of the inner and outer binaries using the equilibrium tide model \citep{1981A&A....99..126H,1998ApJ...499..853E}. When modeling the tidal evolution of the outer binary, we treat the inner binary as a point mass (i.e., the outer binary only evolves due to tides raised on the tertiary by the inner binary, considered as a point mass). Note that outer binary tides are only included for completeness; in our systems, outer binary tides are  negligible. The equilibrium tide model is implemented using equations (81) and (82) of \citet{1998ApJ...499..853E}, with the non-dissipative terms $X$, $Y$ and $Z$ given explicitly by equation (10)-(12) of \citet{2001ApJ...562.1012E}, and the dissipative terms $V$ given explicitly by equations (A7)-(A11) of \citet{2009MNRAS.395.2268B}. The stellar spins are allowed to change in direction and magnitude owing to tidal evolution, although we initialize the spins with zero obliquity. The initial spin period of all stars is set to 10 d. 

The tidal dissipation strength (tidal time lag) is computed as a function of the stellar properties using the prescription of \citet{2002MNRAS.329..897H}. Following \citet{2007ApJ...669.1298F}, we assume a fixed apsidal motion constant of $k_{\mathrm{AM},i}=0.014$, and a gyration radius of 0.08. 

The details of tidal evolution are still highly uncertain (see, e.g., \citealt{2014ARA&A..52..171O} for a review). At high eccentricities, when the orbits are close to parabolic, dynamical tides may provide a better description than the equilibrium tide model (e.g., \citealt{1977ApJ...213..183P,1995ApJ...450..732M}). A hybrid method such as adopted by \citet{2018ApJ...854...44M} is beyond the scope of this paper.

\subsection{Flybys}
\label{sect:meth:flybys}
We model perturbations from passing stars on the triple system in the impulsive approximation. In this case, the orbital motion in the triple system is much slower than the motion of the perturbing star, and the latter imparts a velocity kick on the three stars. This impulsive approximation is justified for wide triples (typical orbits wider than $12\,\au$), for which the typical orbital speeds ($\lesssim 10\,\mathrm{km\,s^{-1}}$), are lower than the typical encounter speed ($\sim 30\,\kms$). We sample perturbing stars from an encounter sphere with a radius of $R_\mathrm{enc}=10^4\,\au$. The local stellar number density is assumed to be $n_\star = 0.1 \, \mathrm{pc^{-3}}$ \citep{2000MNRAS.313..209H}, with a one-dimensional velocity dispersion $\sigma_\star = 30\,\mathrm{km\,s^{-1}}$. The perturbing stars' mass function is assumed to be a Kroupa mass function \citep{1993MNRAS.262..545K} between 0.1 and 80 $\msun$, corrected for gravitational focusing and stellar evolution (assuming a perturber age of 10 Gyr). 

For more details on our methodology to sample perturbing stars and to compute their effect on the inner and outer orbits, we refer to \citet{2017AJ....154..272H,2018MNRAS.476.4139H,2018MNRAS.478..620H,2019MNRAS.482.2262H}.

\subsection{WD kicks}
\label{sect:meth:kick}
Following the results of \citet{2018MNRAS.480.4884E}, we apply an instantaneous kick velocity, $\ve{V}_\kick$, to any star that evolves to a WD. The direction of the kick is assumed to be isotropic, whereas the magnitude is sampled from a Maxwellian distribution with $\sigma_\kick = 0.5 \, \kms$, i.e., the probability for a kick magnitude $V_\kick$ is given by
\begin{align}
P(V_\kick) = \sqrt{ \frac{2}{\pi}} \frac{V_\kick^2}{\sigma_\kick^3} \exp \left ( - \frac{V_\kick^2}{2\sigma_\kick^2} \right ).
\end{align}
This distribution peaks at $V_\kick=\sqrt{2} \sigma_\kick \simeq 0.71\,\kms$, and the mean is $\langle V_\kick \rangle = 2\sqrt{2/\pi} \, \sigma_\kick \simeq 0.80 \,\kms$. Assuming a random orbital phase, we compute the new inner and outer orbits using the methodology of \citet{2018MNRAS.476.4139H}. We stop the simulation if the inner and/or outer orbits are no longer bound after the kick. Given the typically low kick velocities, in reality, there might be strong interactions such as collisions following unbound orbits due to WD kicks. This should be investigated in future study using direct $N$-body integrations (see also \S~\ref{sect:conclusions}).

\subsection{Stopping conditions}
\label{sect:meth:sc}
We impose a number of stopping conditions in our simulations. The stopping conditions are:
\begin{enumerate}
\item Roche lobe overflow (RLOF) occurs in the inner binary system. To evaluate the onset of RLOF, we determine the Roche lobe radius at periapsis using the fits of \citet{2007ApJ...660.1624S}, in particular, their equations~(47) through (52). Note that this approach includes RLOF in eccentric orbits. The RLOF stopping condition is not evaluated for WDs, in which case we check for physical collisions. 

We do not model the subsequent {\it triple} evolution after RLOF has ensued, which is complex particularly for eccentric orbits (see, e.g., \citealt{2019ApJ...872..119H} for work towards modeling mass transfer in eccentric triples). However, we do estimate the outcome of RLOF induced by the tertiary by neglecting the subsequent influence of the tertiary, and evolving the inner binary using the \textsc{BSE} code within \textsc{AMUSE}. In contrast to the prior triple evolution, mass transfer and common-envelope evolution are modeled during the isolated binary evolution with \textsc{BSE}. Here, we assume a common-envelope binding energy factor of $\lambda=0.5$, and a common-envelope efficiency of $\alpha=1$. 
\item The triple system becomes dynamically unstable. Stability is evaluated using the criterion of \citet{2001MNRAS.321..398M}. 
\item The secular approximation made in the equations of motion breaks down. This occurs if the timescale for the inner binary angular momentum to change significantly becomes comparable to the inner or outer orbital periods (e.g., \citealt{2004AJ....128.2518C,2005MNRAS.358.1361I,2012arXiv1211.4584K,2012ApJ...757...27A,2013PhRvL.111f1106S,2014ApJ...781...45A,2014MNRAS.439.1079A,2014MNRAS.438..573B,2016MNRAS.458.3060L,2018MNRAS.481.4907G}). To check for this regime, also known as the semisecular regime, we use the criterion of \citet{2014ApJ...781...45A} which can be derived by equating the timescale for the inner binary angular momentum to change by order itself to the inner orbital period, i.e.,
\begin{align}
\sqrt{1-e_1} < 5 \pi \frac{m_3}{m_1+m_2} \left [ \frac{a_1}{a_2 \left(1-e_2 \right )} \right ]^3.
\end{align}
\item The system becomes unbound, either due to flybys, or due to WD kicks.
\item The stars in the inner binary merge due to high eccentricity (only occurs in the case of WDs). 
\item The system age exceeds 10 Gyr.
\end{enumerate}

\subsection{Simulation setup}
\label{sect:meth:IC}
We carry out four sets of simulations: all combinations of with and without flybys taken into account, and with and without kicks applied to newly-formed WDs. Each set consists of $N_\mathrm{MC}=10^4$ systems which are sampled using a Monte Carlo approach. Our assumptions of the distributions of the initial parameters are described below. An overview is given in the third column of Table~\ref{table:IC}.

\subsubsection{Masses}
\label{sect:meth:IC:mass}
Our interest is in triples in which the components in the inner binary evolve to CO WDs within 10 Gyr (assuming Solar metallicity). Therefore, we restrict to systems with primary mass $1<m_1/\msun<6.5$, and secondary mass $m_2 > 1\, \msun$. The primary mass is sampled from a Kroupa initial mass function \citep{1993MNRAS.262..545K}, i.e., $\mathrm{d}N/\mathrm{d}m_1 \propto m_1^{-2.7}$ for $1<m_1/\msun<6.5$. We assume a flat distribution of the mass ratio $q_1 = m_2/m_1$.

We sample the tertiary mass $m_3$ assuming a flat distribution of the mass ratio $q_2 = m_3/(m_1+m_2)$, with $0.01<q_2<1$ and $m_3>0.1\,\msun$, implying that the tertiary mass is correlated with the inner binary mass.

\subsubsection{Orbits}
\label{sect:meth:IC:orbits}
To sample the orbital semimajor axes, we draw two orbital periods from a Gaussian distribution in $\mathrm{log}_{10}(P_{\mathrm{orb},i}/\mathrm{d})$, with a mean of 5.03 and standard deviation of 2.28, and $0<\mathrm{log}_{10}(P_{\mathrm{orb},i}/\mathrm{d})<10$ \citep{2010ApJS..190....1R}. Two eccentricities are drawn from a Rayleigh distribution between 0.01 and 0.9 with an rms width of 0.33 \citep{2010ApJS..190....1R}. We reject systems that do not satisfy initial dynamical stability using the \citet{2001MNRAS.321..398M} criterion. In addition, we restrict to wide systems, i.e., triples in which the inner binary would not interact in the absence of a tertiary. This is achieved by rejecting systems with $a_1(1-e_1^2) < 12\, \au$. We return to a discussion of this restriction in \S s~\ref{sect:discussion:compact} and \ref{sect:conclusions}.

We assume initially random orientations for both inner and outer orbits. Although observed compact triples ($a_2 \lesssim 100\,\au$) show evidence for more aligned inner and outer orbits (compatible with disk fragmentation), this is a reasonable assumption for wider triples \citep{2017ApJ...844..103T}, which are the focus in this paper.

\section{Results}
\label{sect:results}
We discuss the fractions of the most important outcomes in our simulations in \S\,\ref{sect:results:frac}. We focus on statistical orbital properties and evolution in \S\,\ref{sect:results:orb}. WD mergers and their implications for SNe Ia progenitors are discussed in \S\,\ref{sect:results:col}.

\subsection{Outcome fractions}
\label{sect:results:frac}

\definecolor{Gray}{gray}{0.9}
\begin{table}
\scriptsize
\begin{tabular}{lcccc}
\toprule
& \multicolumn{4}{c}{Fraction} \\
& \multicolumn{2}{c}{No WD kicks} &\multicolumn{2}{c}{WD kicks} \\
& NF & F & NF & F \\
\midrule
No Interaction & $0.806 \pm 0.009$ & $0.767 \pm 0.009$ & $0.086 \pm 0.003$ & $0.166 \pm 0.004$ \\
\midrule
RLOF $\star$1 & $0.107 \pm 0.003$ & $0.112 \pm 0.003$ & $0.099 \pm 0.003$ & $0.104 \pm 0.003$ \\
\quad MS & $0.007 \pm 0.001$ & $0.009 \pm 0.001$ & $0.006 \pm 0.001$ & $0.009 \pm 0.001$ \\
\quad G & $0.090 \pm 0.003$ & $0.093 \pm 0.003$ & $0.084 \pm 0.003$ & $0.086 \pm 0.003$ \\
\quad CHeB & $0.010 \pm 0.001$ & $0.010 \pm 0.001$ & $0.009 \pm 0.001$ & $0.009 \pm 0.001$ \\
\midrule
RLOF $\star$2 & $0.016 \pm 0.001$ & $0.019 \pm 0.001$ & $0.021 \pm 0.001$ & $0.020 \pm 0.001$ \\
\quad MS & $0.007 \pm 0.001$ & $0.007 \pm 0.001$ & $0.009 \pm 0.001$ & $0.009 \pm 0.001$ \\
\quad G & $0.007 \pm 0.001$ & $0.009 \pm 0.001$ & $0.010 \pm 0.001$ & $0.010 \pm 0.001$ \\
\quad CHeB & $0.001 \pm 0.000$ & $0.002 \pm 0.000$ & $0.002 \pm 0.000$ & $0.001 \pm 0.000$ \\
\midrule
Dynamical inst. & $0.038 \pm 0.002$ & $0.056 \pm 0.002$ & $0.164 \pm 0.004$ & $0.178 \pm 0.004$ \\
\quad MS+MS & $0.002 \pm 0.000$ & $0.004 \pm 0.001$ & $0.025 \pm 0.002$ & $0.025 \pm 0.002$ \\
\quad G+MS & $0.009 \pm 0.001$ & $0.009 \pm 0.001$ & $0.012 \pm 0.001$ & $0.011 \pm 0.001$ \\
\quad WD+MS & $0.006 \pm 0.001$ & $0.010 \pm 0.001$ & $0.065 \pm 0.003$ & $0.063 \pm 0.003$ \\
\quad WD+G & $0.015 \pm 0.001$ & $0.013 \pm 0.001$ & $0.015 \pm 0.001$ & $0.017 \pm 0.001$ \\
\quad WD+WD & $0.004 \pm 0.001$ & $0.019 \pm 0.001$ & $0.038 \pm 0.002$ & $0.056 \pm 0.002$ \\
\midrule
Semisecular & $0.022 \pm 0.001$ & $0.029 \pm 0.002$ & $0.021 \pm 0.001$ & $0.023 \pm 0.002$ \\
\quad MS+MS & $0.015 \pm 0.001$ & $0.017 \pm 0.001$ & $0.016 \pm 0.001$ & $0.017 \pm 0.001$ \\
\quad G+MS & $0.000 \pm 0.000$ & $0.000 \pm 0.000$ & $0.000 \pm 0.000$ & $0.000 \pm 0.000$ \\
\quad WD+MS & $0.004 \pm 0.001$ & $0.005 \pm 0.001$ & $0.004 \pm 0.001$ & $0.004 \pm 0.001$ \\
\quad WD+G & $0.001 \pm 0.000$ & $0.001 \pm 0.000$ & $0.000 \pm 0.000$ & $0.000 \pm 0.000$ \\
\quad WD+WD & $0.002 \pm 0.000$ & $0.006 \pm 0.001$ & $0.001 \pm 0.000$ & $0.002 \pm 0.000$ \\
\midrule
Secular collision & $0.000 \pm 0.000$ & $0.000 \pm 0.000$ & $0.000 \pm 0.000$ & $0.001 \pm 0.000$ \\
\quad WD+WD & $0.000 \pm 0.000$ & $0.000 \pm 0.000$ & $0.000 \pm 0.000$ & $0.001 \pm 0.000$ \\
\midrule
Unbound (flyby) & $0.000 \pm 0.000$ & $0.007 \pm 0.001$ & $0.000 \pm 0.000$ & $0.002 \pm 0.000$ \\
\quad MS+MS & $0.000 \pm 0.000$ & $0.001 \pm 0.000$ & $0.000 \pm 0.000$ & $0.001 \pm 0.000$ \\
\quad G+MS & $0.000 \pm 0.000$ & $0.000 \pm 0.000$ & $0.000 \pm 0.000$ & $0.000 \pm 0.000$ \\
\quad WD+MS & $0.000 \pm 0.000$ & $0.001 \pm 0.000$ & $0.000 \pm 0.000$ & $0.000 \pm 0.000$ \\
\quad WD+G & $0.000 \pm 0.000$ & $0.000 \pm 0.000$ & $0.000 \pm 0.000$ & $0.000 \pm 0.000$ \\
\quad WD+WD & $0.000 \pm 0.000$ & $0.005 \pm 0.001$ & $0.000 \pm 0.000$ & $0.001 \pm 0.000$ \\
\midrule
Unbound (kick) & $0.010 \pm 0.001$ & $0.010 \pm 0.001$ & $0.608 \pm 0.008$ & $0.505 \pm 0.007$ \\
\quad MS+MS & $0.010 \pm 0.001$ & $0.010 \pm 0.001$ & $0.218 \pm 0.005$ & $0.174 \pm 0.004$ \\
\quad G+MS & $0.000 \pm 0.000$ & $0.000 \pm 0.000$ & $0.008 \pm 0.001$ & $0.007 \pm 0.001$ \\
\quad WD+MS & $0.000 \pm 0.000$ & $0.000 \pm 0.000$ & $0.186 \pm 0.004$ & $0.144 \pm 0.004$ \\
\quad WD+G & $0.000 \pm 0.000$ & $0.000 \pm 0.000$ & $0.010 \pm 0.001$ & $0.009 \pm 0.001$ \\
\quad WD+WD & $0.000 \pm 0.000$ & $0.000 \pm 0.000$ & $0.159 \pm 0.004$ & $0.149 \pm 0.004$ \\
\bottomrule
\end{tabular}
\caption{ Fractions of outcomes from the population synthesis runs (after 10 Gyr of evolution, or until a stopping condition occurred). Columns 2-3 (4-5): without (with) WD kicks taken into account. The columns `NF' and `F' indicate simulations in which flybys were excluded and included, respectively. The fractions in each column are based on $N_\mathrm{MC}=10^4$ simulations, and are rounded to three decimal places, with errors based on Poisson statistics. In the outcomes, we make additional distinction between the evolutionary phase --  `G' means giant(like), i.e., an HG, RGB or AGB star, whereas CHeB refers to a core helium-burning star. }
\label{table:fractions}
\end{table}

\definecolor{Gray}{gray}{0.9}
\begin{table}
\scriptsize
\begin{tabular}{lcccc}
\toprule
& \multicolumn{4}{c}{Fraction of RLOF systems ($\star 1$ or $\star$ 2)} \\
& \multicolumn{2}{c}{No WD kicks} &\multicolumn{2}{c}{WD kicks} \\
& NF & F & NF & F \\
\midrule
RLOF $\rightarrow$ merger & $0.380 \pm 0.018$ & $0.374 \pm 0.017$ & $0.398 \pm 0.018$ & $0.401 \pm 0.018$ \\
\quad MS+MS & $0.087 \pm 0.008$ & $0.092 \pm 0.008$ & $0.085 \pm 0.008$ & $0.100 \pm 0.009$ \\
\quad CHeB+MS & $0.013 \pm 0.003$ & $0.009 \pm 0.003$ & $0.015 \pm 0.004$ & $0.010 \pm 0.003$ \\
\quad CHeB+G & $0.013 \pm 0.003$ & $0.007 \pm 0.002$ & $0.013 \pm 0.003$ & $0.008 \pm 0.003$ \\
\quad G+MS & $0.072 \pm 0.008$ & $0.070 \pm 0.007$ & $0.075 \pm 0.008$ & $0.065 \pm 0.007$ \\
\quad G+CHeB & $0.000 \pm 0.000$ & $0.000 \pm 0.000$ & $0.000 \pm 0.000$ & $0.000 \pm 0.000$ \\
\quad G+G & $0.007 \pm 0.002$ & $0.009 \pm 0.003$ & $0.006 \pm 0.002$ & $0.011 \pm 0.003$ \\
\quad He+MS & $0.002 \pm 0.001$ & $0.002 \pm 0.001$ & $0.002 \pm 0.001$ & $0.002 \pm 0.001$ \\
\quad He+G & $0.000 \pm 0.000$ & $0.000 \pm 0.000$ & $0.000 \pm 0.000$ & $0.000 \pm 0.000$ \\
\quad WD+MS & $0.037 \pm 0.006$ & $0.034 \pm 0.005$ & $0.042 \pm 0.006$ & $0.041 \pm 0.006$ \\
\quad WD+G & $0.087 \pm 0.008$ & $0.086 \pm 0.008$ & $0.097 \pm 0.009$ & $0.089 \pm 0.008$ \\
\quad WD+He & $0.015 \pm 0.004$ & $0.019 \pm 0.004$ & $0.012 \pm 0.003$ & $0.020 \pm 0.004$ \\
\quad He WD+WD & $0.037 \pm 0.005$ & $0.036 \pm 0.005$ & $0.037 \pm 0.006$ & $0.039 \pm 0.006$ \\
\quad CO WD+WD & $0.003 \pm 0.002$ & $0.005 \pm 0.002$ & $0.007 \pm 0.002$ & $0.006 \pm 0.002$ \\
\quad ONe WD+WD & $0.000 \pm 0.000$ & $0.000 \pm 0.000$ & $0.000 \pm 0.000$ & $0.000 \pm 0.000$ \\
\quad Other & $0.007 \pm 0.002$ & $0.005 \pm 0.002$ & $0.007 \pm 0.002$ & $0.009 \pm 0.003$ \\
\bottomrule
\end{tabular}
\caption{  Various merger outcome fractions after RLOF was induced by the tertiary. These fractions are with respect to the RLOF systems, which comprise $\sim0.1$ of all systems (see Table~\ref{table:fractions}). Fractions were obtained by evolving the inner binaries of the systems in which RLOF occurred using the \textsc{BSE} code (see also \S~\ref{sect:meth:sc}). Here, `He' refers to a naked helium star. }
\label{table:fractions_RLOF}
\end{table}

\begin{figure}
\center
\includegraphics[scale = 0.47, trim = 5mm -5mm 0mm 10mm]{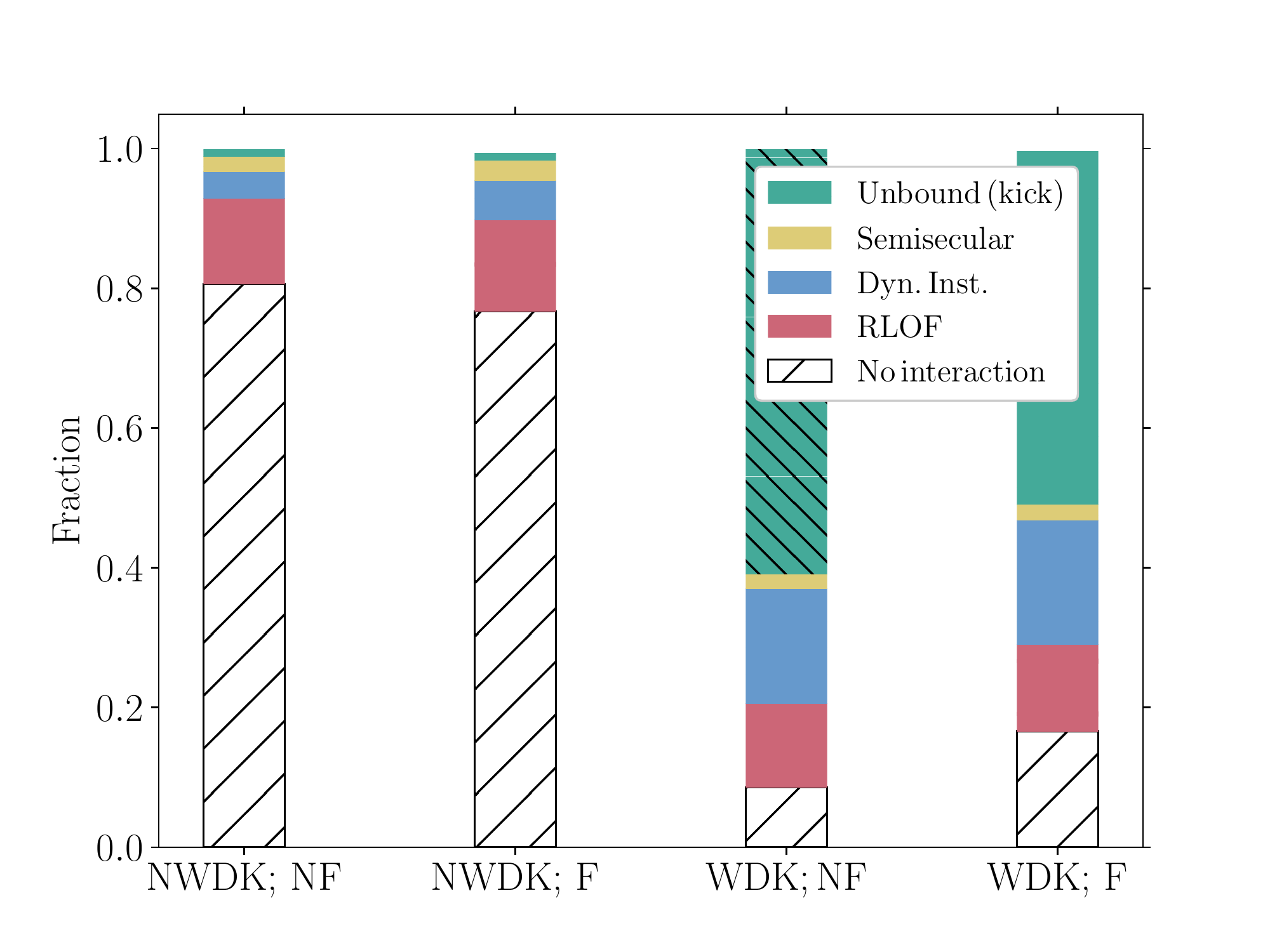}
\caption{Visual representation of the outcome fractions of the main channels in the population synthesis simulations. The four bars correspond to the four data columns of Table~\ref{table:fractions}, i.e., `(N)WDK' refers to (no) WD kicks included, and `(N)F' refers to flybys (not) included. We include the five most important channels (refer to the legend). The `RLOF' channel includes both RLOF from the primary star, as well as from the secondary star. }
\label{fig:bar_plot}
\end{figure}

For the four sets of simulations, each consisting of $N_\mathrm{MC}=10^4$ systems, Table~\ref{table:fractions} enumerates the fractions of several outcomes associated with the stopping conditions (\S\,\ref{sect:meth:sc}). A visual representation of the main channels is given in a bar plot in \F~\ref{fig:bar_plot}. 

If WD kicks are not taken into account (columns 2 and 3), the majority of systems ($\sim 0.8$) do not interact during the evolution, i.e., no RLOF or merger occurs, and the system remains stable. With the inclusion of WD kicks (columns 4 and 5), however, only $\sim $ 0.1-0.2 of systems do not interact.  Instead, a large fraction of systems becomes unbound ($\sim$ 0.5-0.6), or dynamically unstable ($\sim 0.2$).

Flybys (the columns labeled `NF' and `F' correspond to flybys excluded and included, respectively) slightly increase the RLOF, dynamical instability and semisecular regime fractions. This can be understood from potential increased mutual inclination and/or outer orbit eccentricity induced by passing stars, which can subsequently boost LK oscillations. 

In Table~\ref{table:fractions_RLOF}, we show various merger outcome fractions after RLOF was induced by the tertiary. These fractions are with respect to the RLOF systems, which comprise $\sim0.1$ of all systems (see Table~\ref{table:fractions}). The fractions in Table~\ref{table:fractions_RLOF} were obtained by evolving the inner binaries of the systems in which RLOF occurred (potentially in an eccentric orbit, see also \S~\ref{sect:results:orb:e} below) using the \textsc{BSE} code (see also \S~\ref{sect:meth:sc}).

Typically, $\sim0.4$ of systems in which RLOF is triggered in the inner binary, ultimately merge. Most of these mergers occur on the MS. A fraction of $\sim 0.04$ of RLOF systems ($\sim 0.004$ of all systems) result in WD+WD mergers, with the majority comprising the merger of a CO WD with a He WD. 

\subsection{Orbital properties and evolution}
\label{sect:results:orb}

\begin{figure*}
\center
\includegraphics[scale = 0.45, trim = 10mm -5mm 0mm 5mm]{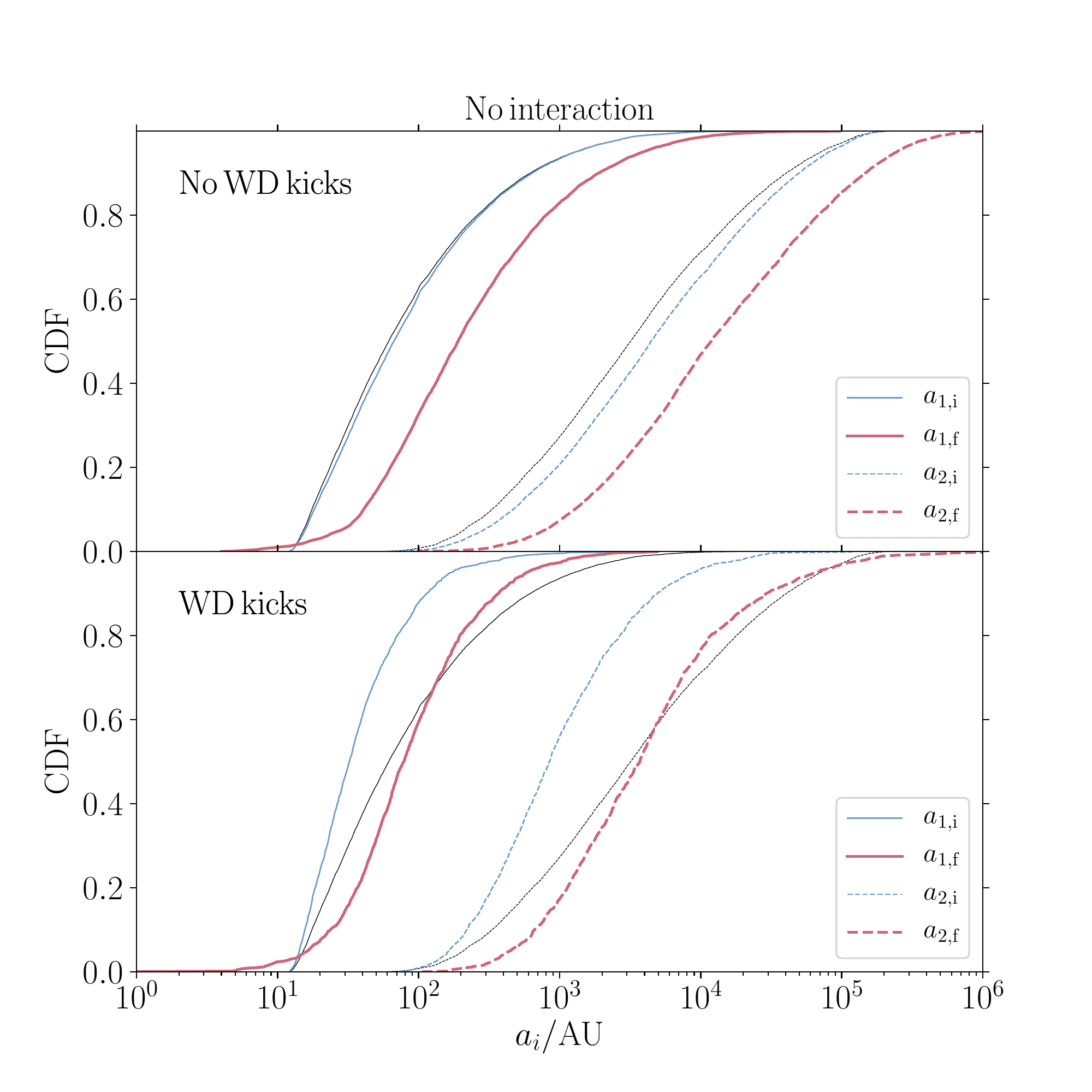}
\includegraphics[scale = 0.45, trim = 10mm -5mm 0mm 10mm]{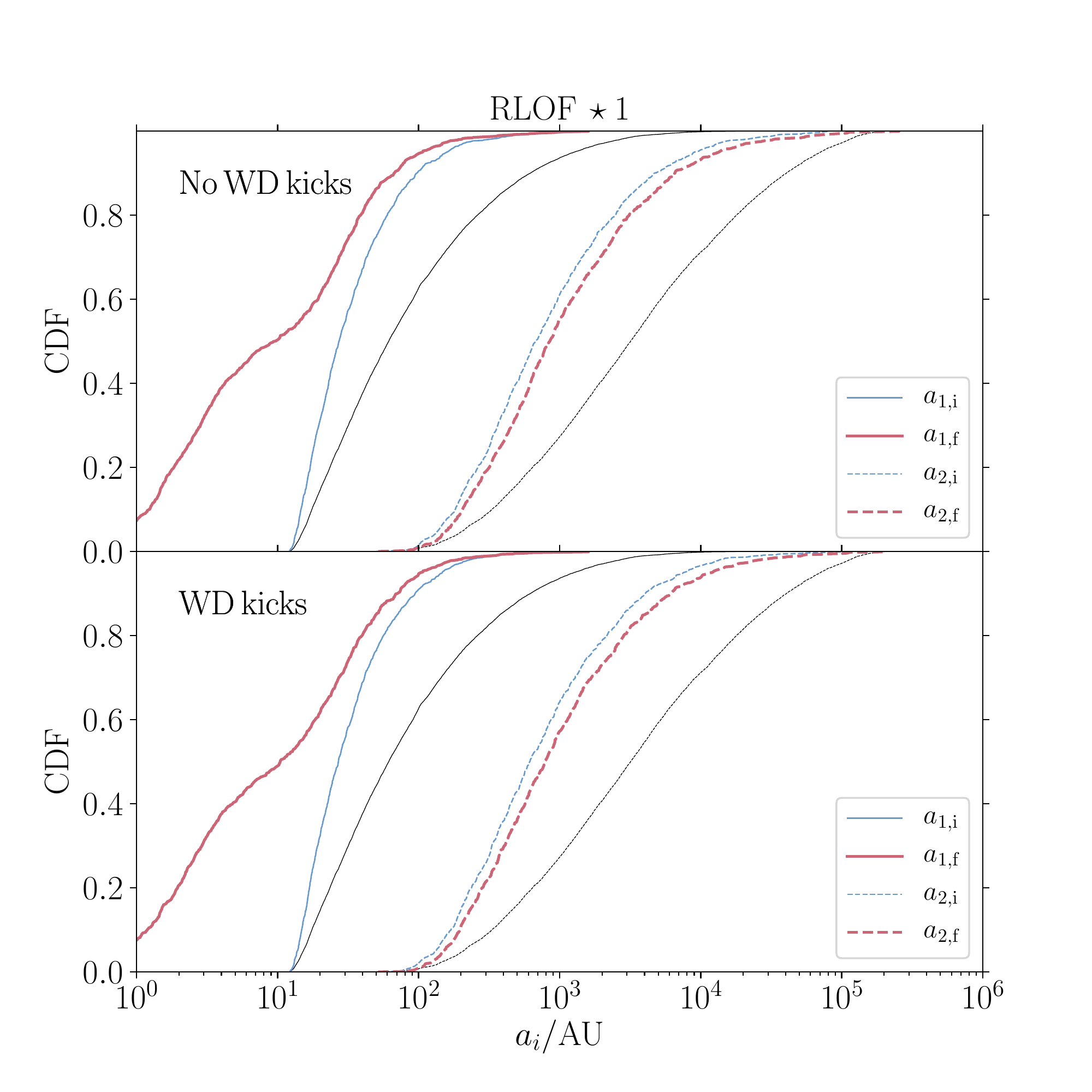}
\includegraphics[scale = 0.45, trim = 10mm -5mm 0mm 5mm]{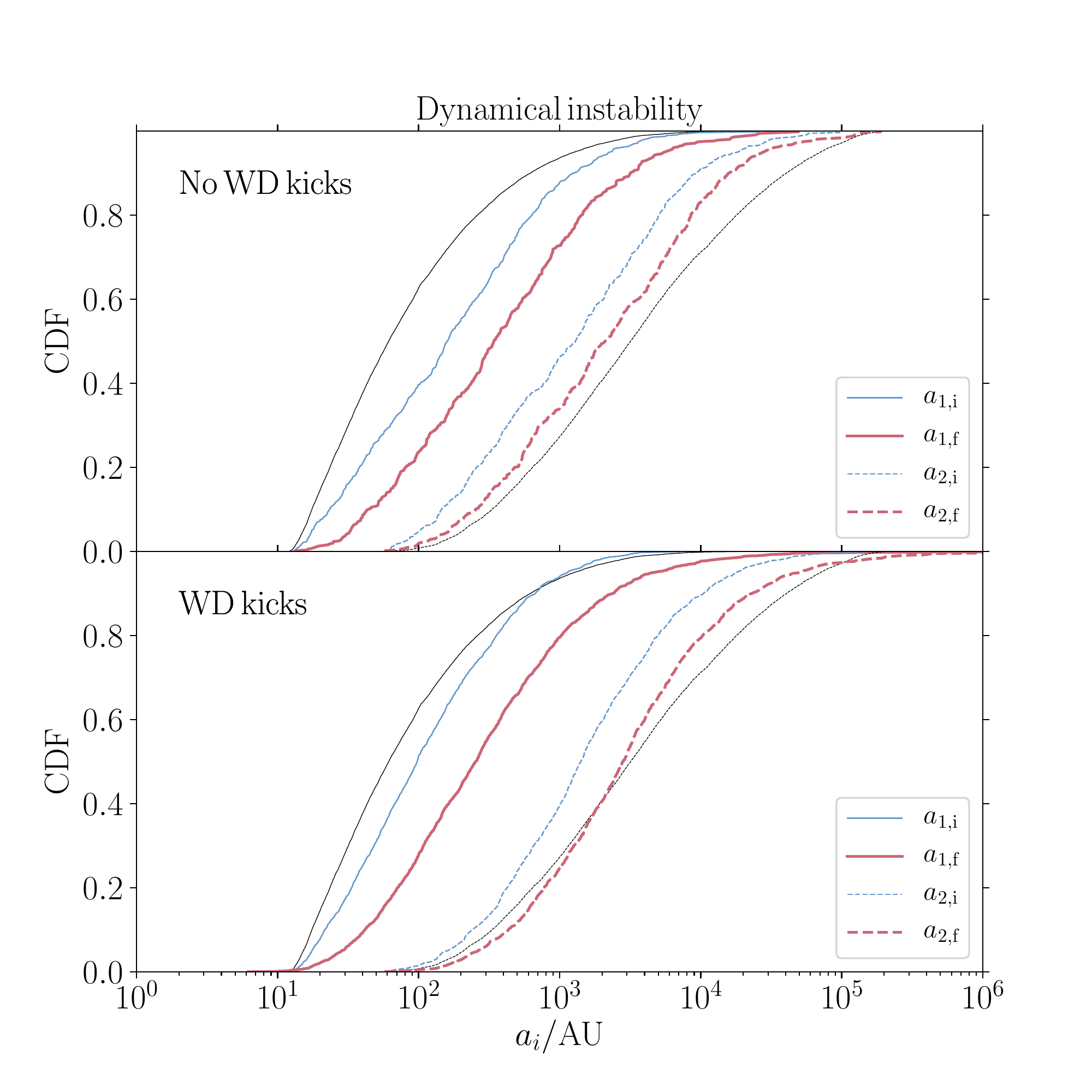}
\includegraphics[scale = 0.45, trim = 10mm -5mm 0mm 10mm]{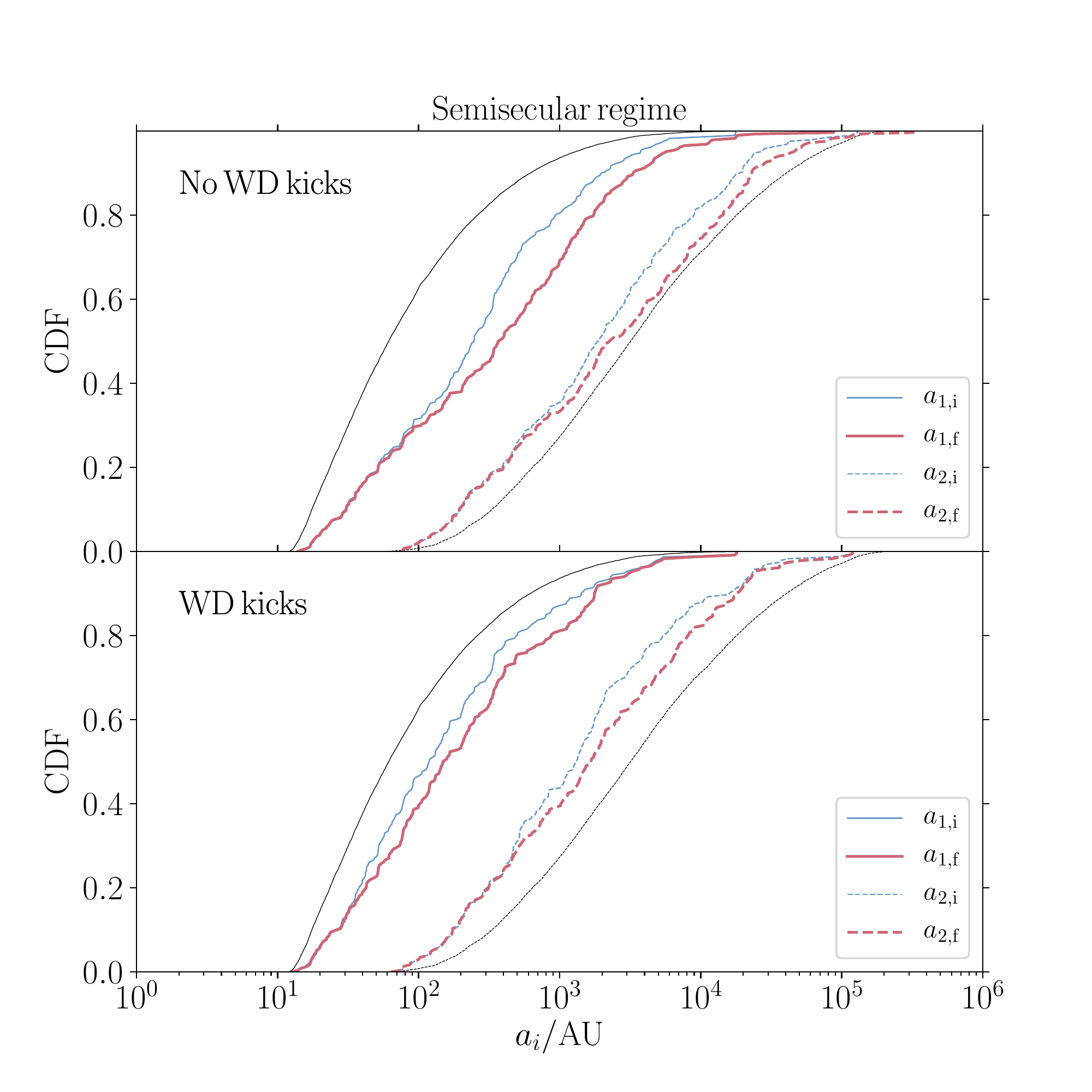}
\caption { Cumulative distributions of the semimajor axes (solid: inner orbit; dashed: outer orbit) corresponding to four different outcomes: no-interaction, RLOF star 1, dynamical instability, and the semisecular regime. Flybys are included in all panels. Thin blue (thick red) lines correspond to the initial (final) distributions. In each set of panels, kicks are excluded and included in the the top and bottom panels, respectively. For reference, the cumulative distributions of the inner and outer orbital semimajor axes of all systems are shown with the think black solid and dotted lines, respectively.  }
\label{fig:pop_syn_sma}
\end{figure*}

\subsubsection{Semimajor axes}
\label{sect:results:orb:sma}
We show cumulative distributions of the semimajor axes (solid: inner orbit; dashed: outer orbit) in \F~\ref{fig:pop_syn_sma} corresponding to four different outcomes: no-interaction, RLOF star 1, dynamical instability, and the semisecular regime. Thin blue (thick red) lines correspond to the initial (final) distributions. In each set of panels, kicks are excluded and included in the the top and bottom panels, respectively. For reference, the cumulative distributions of the inner and outer orbital semimajor axes of all systems are shown with the thin black solid and dotted lines, respectively.

As can be expected, non-interacting systems tend to have relatively wide outer orbits. Both the inner and outer orbits tend to expand due to mass loss (compare the thin blue and thick red lines in \F~\ref{fig:pop_syn_sma}). In some cases, the inner orbit shrinks owing to tidal dissipation triggered by LK oscillations. 

RLOF (star 1) tends to occur in the initially more compact inner systems. By the time RLOF occurs, the semimajor axis has typically shrunk due to tides. A large fraction of systems enter RLOF with eccentric inner orbits (see \S\,\ref{sect:results:orb:e}). 

Triples becoming dynamically unstable or entering the semisecular regime tend to be more compact (smaller $a_2/a_1$ ratio) compared to all systems. The inner and outer orbits expand due to mass loss for both cases of dynamical instability and the semisecular regime. However, orbital expansion is less significant in the case of the semisecular regime -- the latter is mainly driven by LK-induced high eccentricities.

Generally, systems tend to have tighter orbits when WD kicks are included. This can be understood by noting that systems with initially wider orbits tend to be disrupted due to WD kicks.

\begin{figure*}
\center
\includegraphics[scale = 0.45, trim = 10mm -5mm 0mm 5mm]{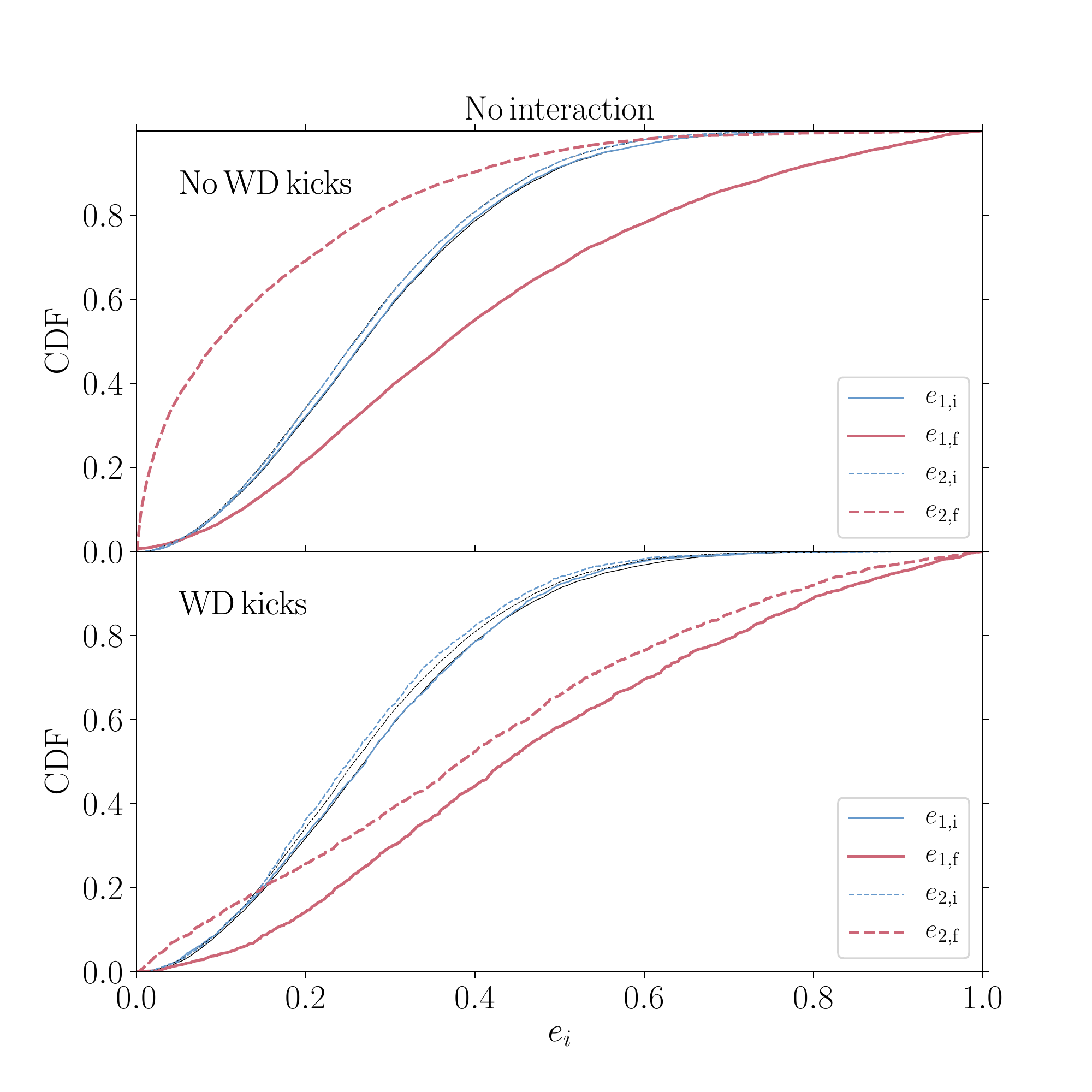}
\includegraphics[scale = 0.45, trim = 10mm -5mm 0mm 10mm]{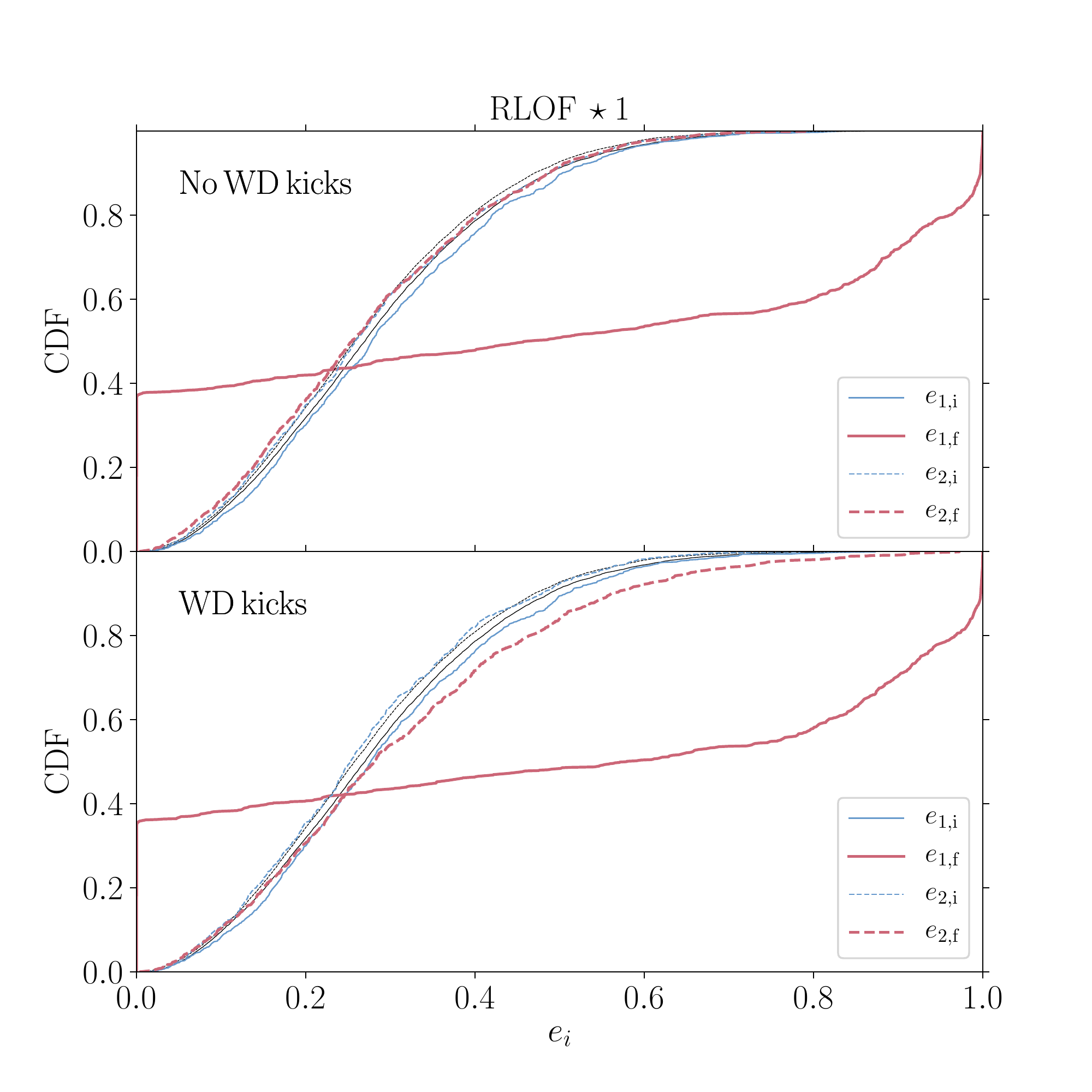}
\includegraphics[scale = 0.45, trim = 10mm -5mm 0mm 5mm]{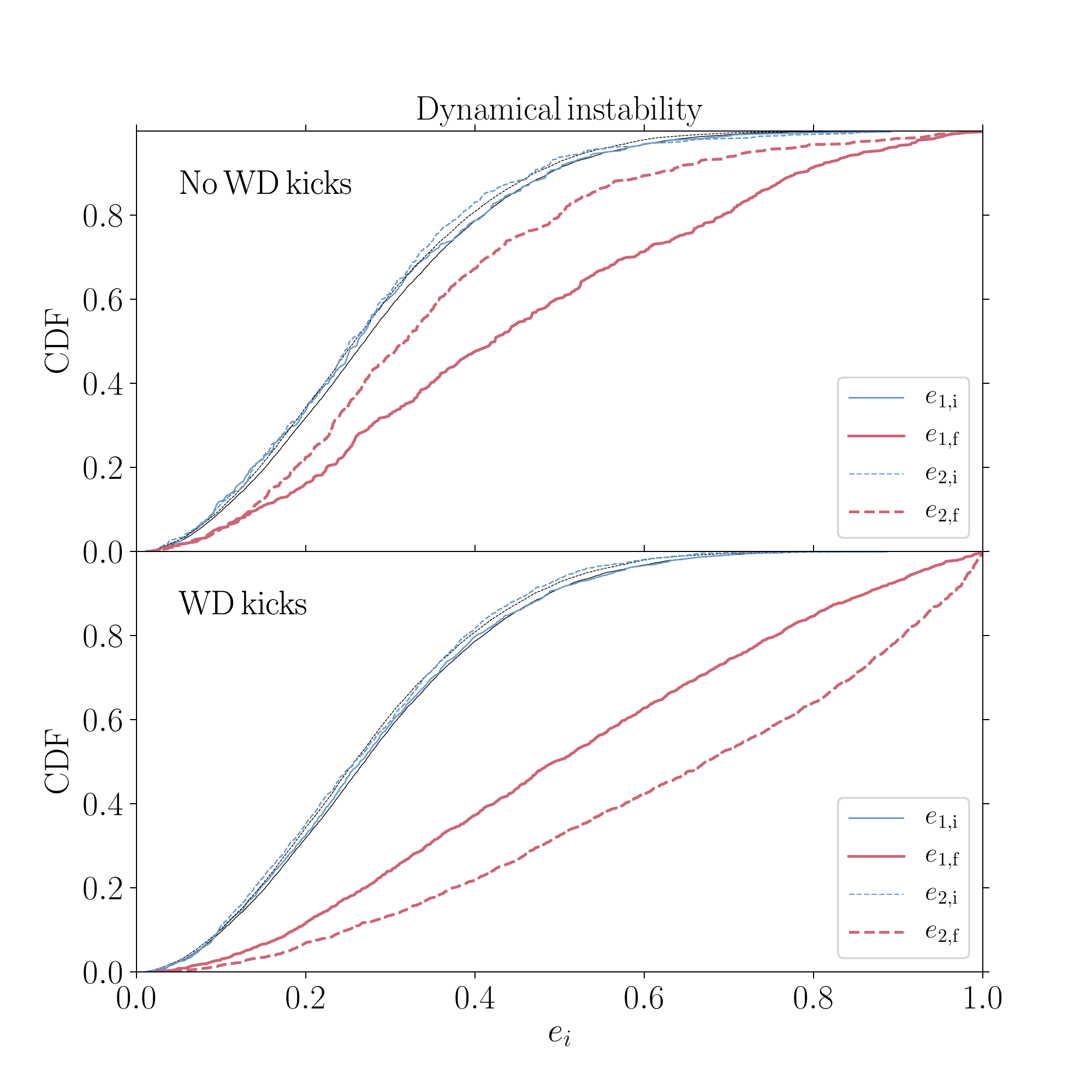}
\includegraphics[scale = 0.45, trim = 10mm -5mm 0mm 10mm]{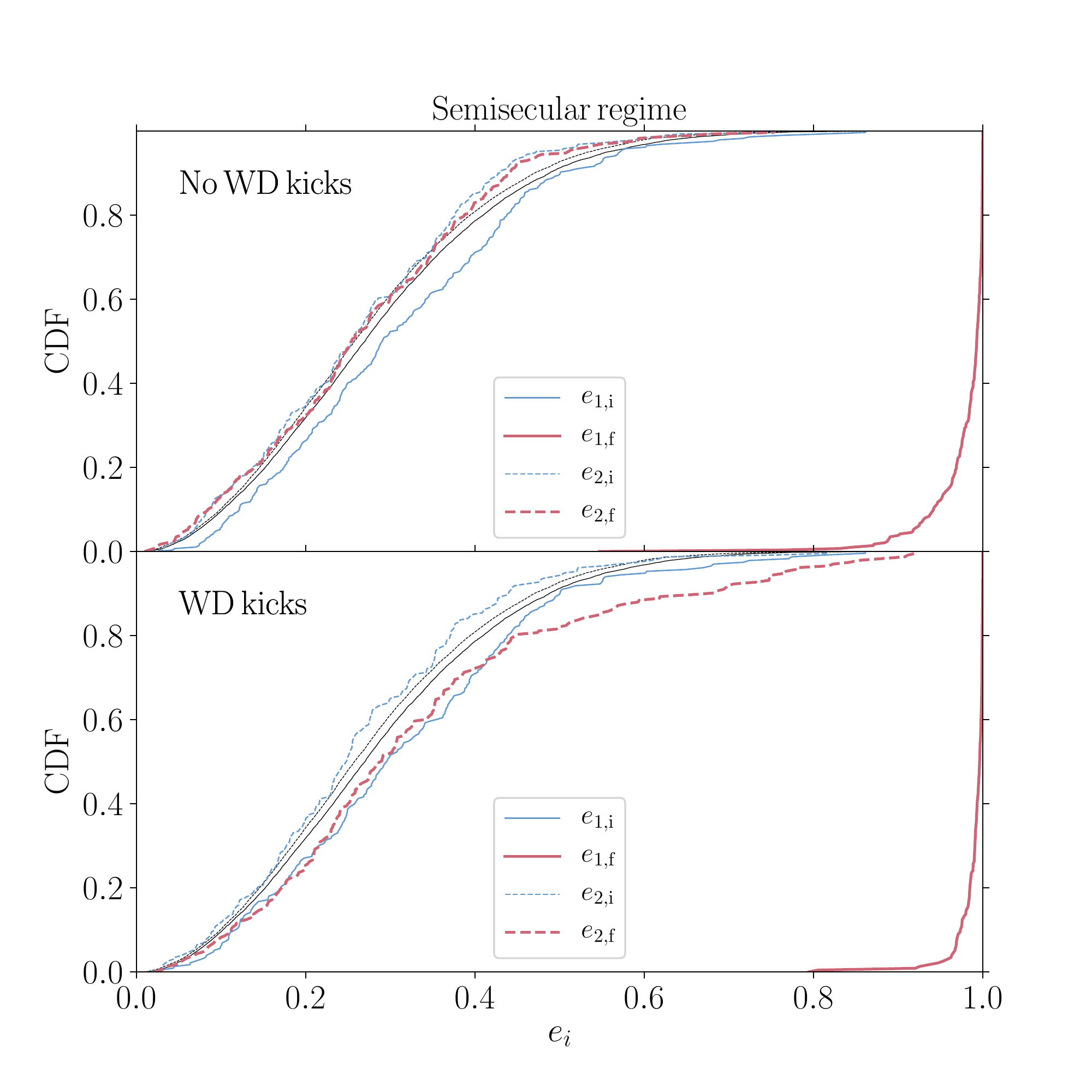}
\caption {Cumulative distributions of the eccentricities for four outcome in the population synthesis simulations similar to \F~\ref{fig:pop_syn_sma}. }
\label{fig:pop_syn_e}
\end{figure*}

\subsubsection{Eccentricities}
\label{sect:results:orb:e}
\F~\ref{fig:pop_syn_e} shows cumulative distributions of the inner (solid lines) and outer (dashed lines) eccentricities, similarly to \F~\ref{fig:pop_syn_sma}. In the non-interacting systems, the inner orbit eccentricity is excited by secular evolution, whereas the outer eccentricity is affected by flybys and/or WD kicks. 

In $\sim 40\%$ of the cases, RLOF in star 1 occurs in circular orbits, whereas for $\sim 30\%$ of cases, $e_{1,\mathrm{f}} > 0.9$. For star 2, RLOF occurs in circular orbits for $\sim 30\%$ of cases, and $e_{1,\mathrm{f}} > 0.9$ for $50\%$ of cases. This illustrates the importance of eccentric RLOF triggered by secular evolution in triple systems (similar proportions of eccentric RLOF apply to quadruple systems, see \citealt{2018MNRAS.478..620H}). 

Systems that become dynamically unstable tend to have higher outer eccentricities, which is easily understood from the dependence of $e_2$ in the stability criterion of \citet{2001MNRAS.321..398M}. By necessity, the inner orbit tends to be highly eccentric ($e_{1,\mathrm{f}}>0.9$ in nearly all cases) for triples to enter the semisecular regime. 

For all four outcomes shown in \F~\ref{fig:pop_syn_e}, WD kicks tend to significantly increase the final outer orbit eccentricity. 

\begin{figure}
\center
\includegraphics[scale = 0.47, trim = 5mm -5mm 0mm 5mm]{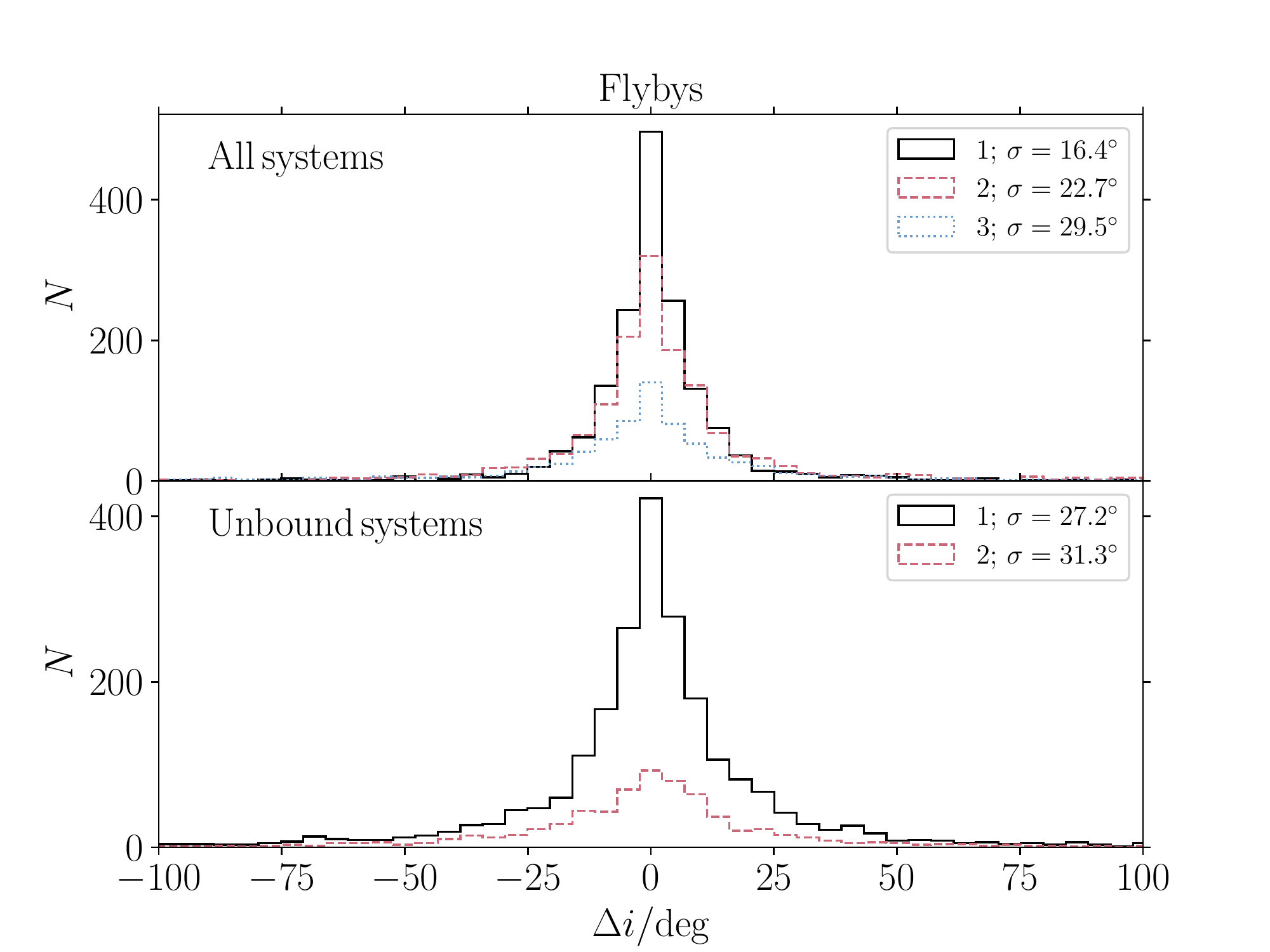}
\caption{ Distributions of the changes in the mutual inclination, $\Delta i$ after a WD kick was imparted when one of the stars evolved to a WD. Black solid lines: after the first WD kick; red dashed lines: after a second WD kick; blue dotted lines: after a third WD kick. Top panels: distributions for all systems in which WD kicks occurred; bottom panel: restricting to those systems that became unbound due to a WD kick. The standard deviations of the distributions are shown in the legend. }
\label{fig:pop_syn_kick_delta_is}
\end{figure}

\begin{figure}
\center
\includegraphics[scale = 0.47, trim = 5mm -5mm 0mm 5mm]{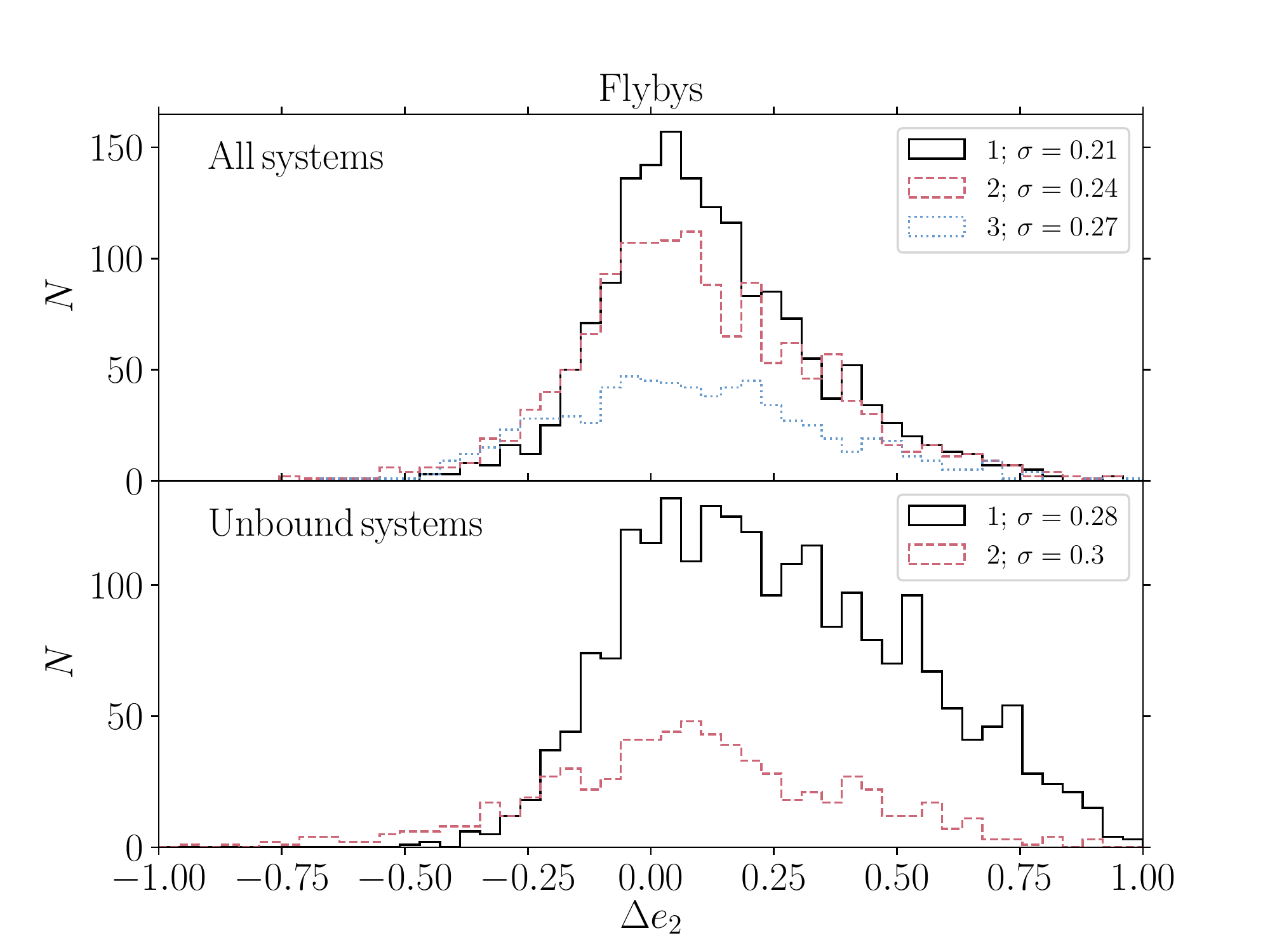}
\caption{Similar to \F~\ref{fig:pop_syn_kick_delta_is}, here showing the distributions of the change of the outer orbit eccentricity. }
\label{fig:pop_syn_kick_delta_es}
\end{figure}

\begin{figure}
\center
\includegraphics[scale = 0.47, trim = 5mm -5mm 0mm 5mm]{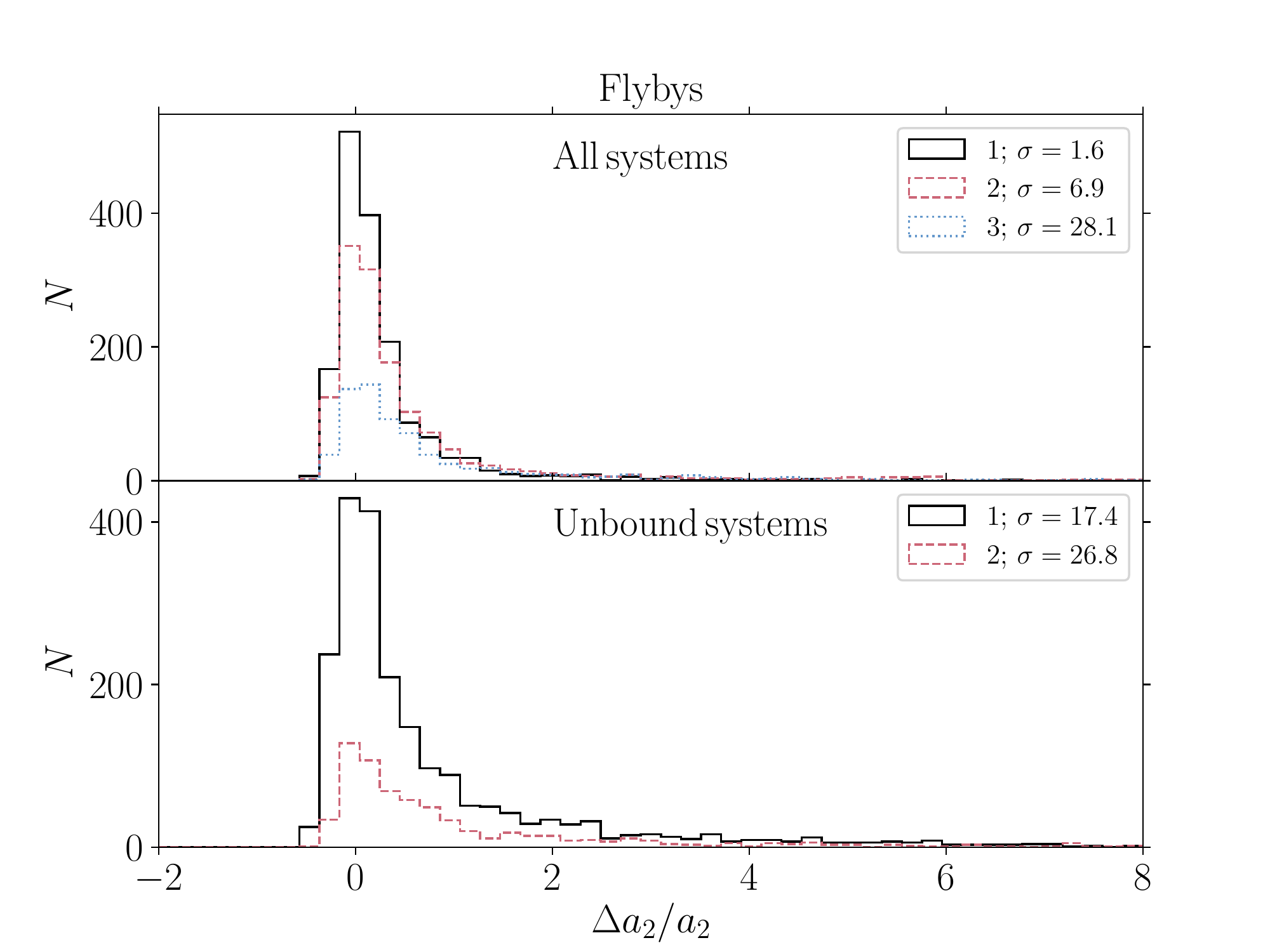}
\caption{Similar to \F~\ref{fig:pop_syn_kick_delta_is}, here showing the distributions of the fractional change of the outer orbit semimajor axis. }
\label{fig:pop_syn_kick_delta_as}
\end{figure}

\subsubsection{The impact of WD kicks}
\label{sect:results:orb:SNe}
Here, we further illustrate how WD kicks impact the orbital evolution of the triple. Specifically, in \F~\ref{fig:pop_syn_kick_delta_is}, \F~\ref{fig:pop_syn_kick_delta_es} and \F~\ref{fig:pop_syn_kick_delta_as}, we show the distributions of the changes in the mutual inclination, $\Delta i$, the outer orbit eccentricity, $\Delta e_2$, and the fractional change in the outer orbit semimajor axis, $\Delta a_2/a_2$, respectively, after a WD kick was imparted when one of the stars evolved to a WD. We include only the cases in which the triple remained bound directly after the WD kick, and make a distinction between the changes after the first WD kick (black solid lines), after a second WD kick (red dashed lines), and after a third (blue dotted lines). In the top panels, we show the distributions for all systems in which WD kicks occurred, whereas in the bottom panels, we restrict to those systems that ultimately became unbound due to a WD kick. In the latter case, there can clearly only be at most two WD kicks for which the system remained bound. 

The distributions of $\Delta i$ are symmetric around $\Delta i = 0^\circ$ (see \F~\ref{fig:pop_syn_kick_delta_is}). The first WD kick results in an inclination change of typically $\simeq 16^\circ$. Later kicks have a larger effect, which can be understood from the fact that orbits have typically expanded after further evolution, such that the effects of WD kicks are larger. The distribution of $\Delta e_2$ (see \F~\ref{fig:pop_syn_kick_delta_es}) is not symmetric; there is a tendency for $e_2$ to increase. The typical change is $\Delta e_2 \sim 0.2$. Similarly, the fractional semimajor axis change (see \F~\ref{fig:pop_syn_kick_delta_as}) is asymmetrically distributed and typically positive, with a typical value of $\sim 1.5$ (note that the standard deviation in $\Delta a_2/a_2$ in \F~\ref{fig:pop_syn_kick_delta_as} is skewed to high values due to a small number of outliers).

\subsection{SNe Ia from WD mergers}
\label{sect:results:col}

\subsubsection{Channels}
\label{sect:results:col:chan}
We consider a number of channels in which mergers of WDs could result in SNe Ia. In particular, we consider collisions of CO WDs, which are likely to result in SNe Ia (e.g., \citealt{2009MNRAS.399L.156R,2009ApJ...705L.128R,2010ApJ...724..111R,2012ApJ...747L..10P,2015ApJ...807..105S}). We also consider SNe Ia from WD mergers in circular orbits driven by gravitational wave emission, which originated from tertiary-induced RLOF. Our channels are:

\begin{enumerate}
\item {\it Secular collisions}: direct collisions of two CO WDs owing to extremely high eccentricities induced by secular evolution. Throughout the evolution, the system did not enter the semisecular regime. This scenario occurs only extremely rarely (see Table~\ref{table:fractions}). 
\item {\it Dynamical instability}: collisions of two CO WDs resulting from a dynamical instability involving at least two WDs. A collision following a dynamical instability is certainly not guaranteed; to determine in which systems dynamical instability can lead to WD collisions, direct $N$-body simulations would have to be carried out. Instead, we here simply assume a certain collision efficiency following dynamical instability. Using direct $N$-body simulations, \citet{2012ApJ...760...99P} found collision probabilities of about 10\%. However, the latter number is based on the entire triple stellar evolution, including the AGB phase during which the stars are much larger, giving a much larger collision cross section. The collision probability following dynamical instability of WD systems is likely lower, and merits detailed investigation. Nevertheless, we here take an optimistic approach and assume that 10\% of our systems that become dynamically unstable with two WDs in the inner binary lead to WD collisions. This is a significant uncertainty that we return to in \S~\ref{sect:conclusions}.
\item {\it Semisecular regime}: WDs can collide after entering the semisecular regime. The delay time can be estimated as \citep{2012arXiv1211.4584K}
\begin{align}
\label{eq:t_col}
\overline{t}_{\mathrm{col}} = \frac{a_1}{4 R_\mathrm{WD}} P_{\mathrm{orb},1},
\end{align}
where $R_\mathrm{WD}$ is the WD radius. Once a triple enters the semisecular regime, we assume that WD collision occurs with a delay time given by at a time given by $\overline{t}_{\mathrm{col}} $. 
\item {\it RLOF-induced circular mergers}: as described in \S~\ref{sect:meth:sc}, in our simulations, if the tertiary induced RLOF in the inner binary, the latter was continued to be evolved in isolation and including binary interaction (mass transfer and common-envelope evolution) using \textsc{BSE}. After one or more phases of mass transfer and/or common-envelope evolution, some of these systems become compact WD systems that ultimately merge due to gravitational wave emission. We find both CO WD+He WD and CO WD+CO WD mergers from this channel, approximately in the ratio $10:1$ (see Table~\ref{table:fractions_RLOF}). Here, we assume that both types of WD mergers result in SNe Ia, although this may not be the case for all these systems depending on the masses (see, e.g., \citealt{2012ApJ...747L..10P}). 
\end{enumerate}

\subsubsection{Normalization}
\label{sect:results:col:norm}
The SNe Ia rates are normalized by computing the total mass $M_\mathrm{tot}$ of the population represented by the $N_\mathrm{calc}=10^4$ calculated systems (this approach is similar to that of \citealt{2013MNRAS.430.2262H}). We consider a galactic population of $N_\mathrm{tot}$ gravitationally bound systems with $N_\mathrm{bin} = \alpha_\mathrm{bin} N_\mathrm{tot}$ binaries, $N_\mathrm{tr} = \alpha_\mathrm{tr} N_\mathrm{tot}$ triples, and $(1-\alpha_\mathrm{bin} - \alpha_\mathrm{tr}) N_\mathrm{tot}$ remaining single stars. We neglect the contribution from quadruple stars (see \citealt{2018MNRAS.478..620H} for a study of the latter). The assumed multiplicity fractions are $\alpha_\mathrm{bin}=0.6$ and $\alpha_\mathrm{tr} = 0.25$ \citep{2010ApJS..190....1R}.

The computed triples represent a fraction, $f_\mathrm{calc}$, of all $N_\mathrm{tr}$ triple systems. We determine $f_\mathrm{calc}$ by sampling systems using the same general assumptions as in the Monte Carlo simulations, except now including all triple systems ($0.1<m_1/\msun<80$, $m_2>0.1\,\msun$, and  $0<\mathrm{log}_{10}[P_{\mathrm{orb},i}/\mathrm{d}]<10$). The fraction $f_\mathrm{calc}$ is then the fraction of systems satisfying the conditions of the systems included in the Monte Carlo simulations ($1<m_1/\msun<6.5$,  $m_2 > 1\,\msun$, and $a_1[1-e_1^2] > 12 \, \au$), yielding $f_\mathrm{calc} \simeq 0.027$. 

We estimate the total mass of all single, binary and triple systems as follows. For a Kroupa mass distribution, the average mass is $M_\mathrm{Kr} \simeq 0.5006\,\msun$ (see, e.g., equation A4 of \citealt{2013MNRAS.430.2262H}). Assuming flat mass ratio distributions, the mass in single stars is $M_\mathrm{s} \approx M_\mathrm{Kr} N_\mathrm{s}$. The total mass of the binary stars is $M_\mathrm{bin} \approx \left(1+\frac{1}{2} \right)M_\mathrm{Kr} N_\mathrm{bin} = \frac{3}{2} M_\mathrm{Kr} N_\mathrm{bin}$, and for the triples, the total mass is $M_\mathrm{tr} \approx \left(\frac{3}{2}+\frac{1}{2} \frac{3}{2}\right)M_\mathrm{Kr} N_\mathrm{tr} = \frac{9}{4} M_\mathrm{Kr} N_\mathrm{tr}$. Addition of all these masses gives
\begin{align}
\nonumber M_{\mathrm{tot}} &\approx M_\mathrm{Kr} \left (N_\mathrm{s} + \frac{3}{2} N_\mathrm{bin} + \frac{9}{4} N_\mathrm{tr} \right ) \\
&= M_\mathrm{Kr} \frac{N_\mathrm{calc}}{f_{\mathrm{calc}} \alpha_\mathrm{tr}} \left (1 + \frac{1}{2} \alpha_\mathrm{bin} + \frac{5}{4} \alpha_\mathrm{tr}  \right ).
\end{align}
With our adopted numbers, $M_\mathrm{tot} \approx 1.2 \times 10^6\,\msun$. We use this total mass to normalize our SNe Ia rates presented below in \S\,\ref{sect:results:col:DTD}.

\begin{figure}
\center
\includegraphics[scale = 0.47, trim = 5mm 0mm 0mm 0mm]{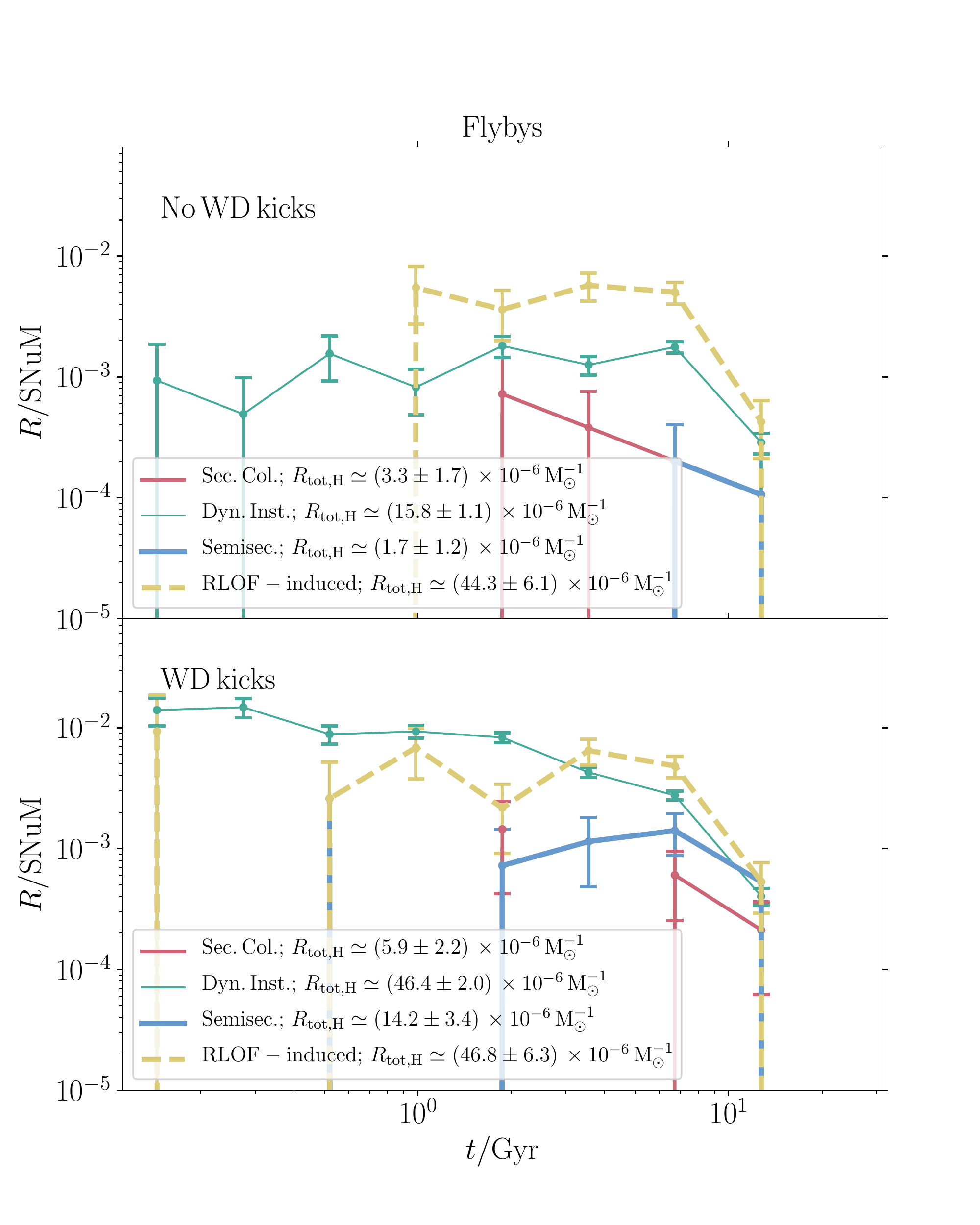}
\caption { DTDs for the individual CO WD merger-induced SNe Ia channels described in \S\,\ref{sect:results:col:chan}, expressed in units of SNuM, i.e., rate per century per $10^{10}\,\msun$ of solar mass. Flybys are included. The time-integrated rates are given in the legends, with the errors based on Poisson statistics. }
\label{fig:DTD_ind}
\end{figure}

\begin{figure}
\center
\includegraphics[scale = 0.48, trim = 5mm 0mm 0mm 0mm]{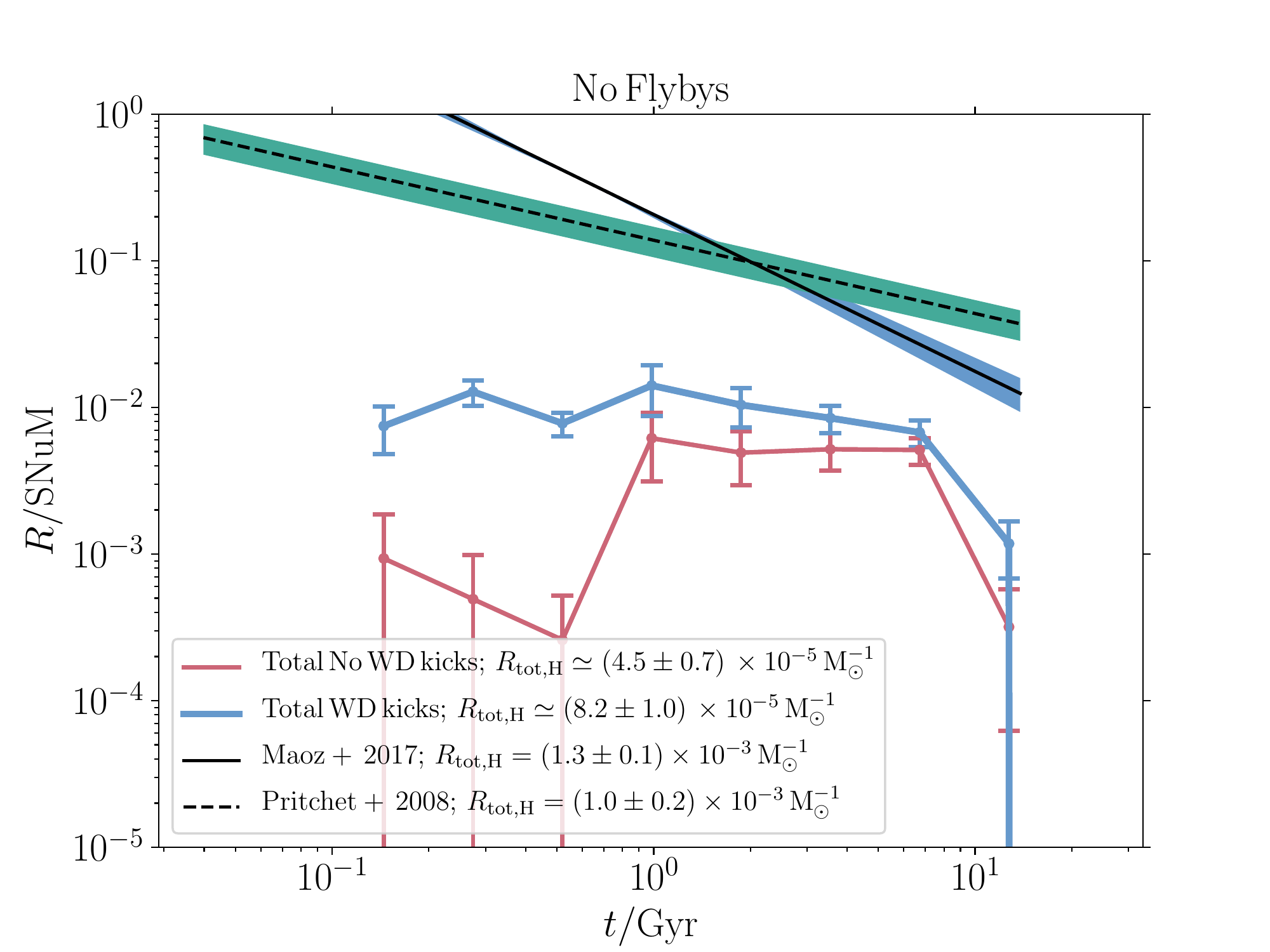}
\includegraphics[scale = 0.48, trim = 5mm 0mm 0mm 0mm]{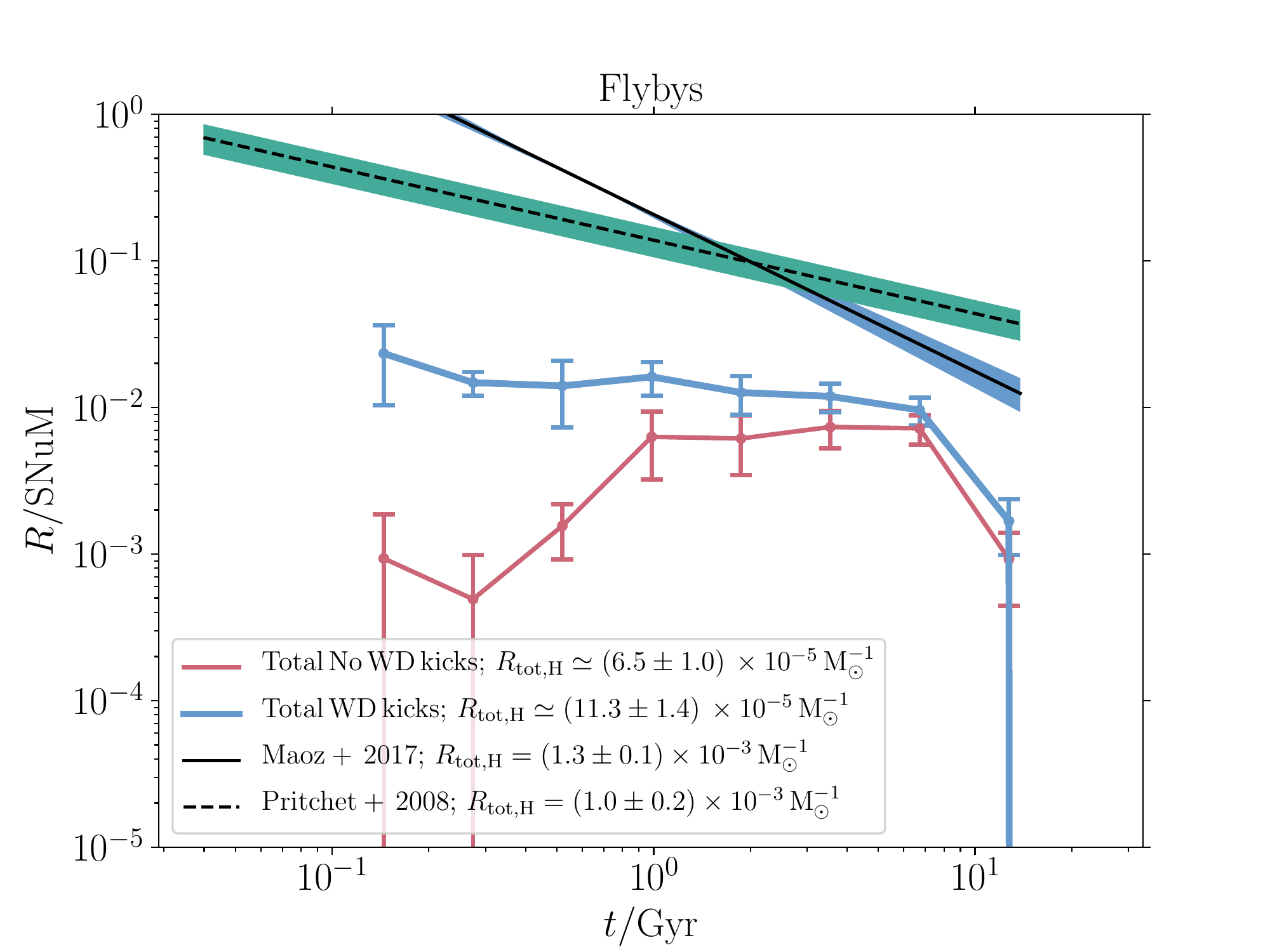}
\caption { DTDs for the combined CO WD merger-induced SNe Ia channels described in \S\,\ref{sect:results:col:chan}, expressed in units of SNuM, i.e., rate per century per $10^{10}\,\msun$ of solar mass. Top and bottom panels: flybys excluded and included, respectively. Thick blue (thin red) lines show the DTDs with WD kicks included (excluded). The time-integrated rates are given in the legends, with the errors based on Poisson statistics. Black solid lines with blue error regions show the observed DTD from \citet{2017ApJ...848...25M}. The dashed black lines with green error regions show the DTD from \citeauthor{2008ApJ...683L..25P} (\citeyear{2008ApJ...683L..25P}. For both the DTDs of \citet{2017ApJ...848...25M} and \citet{2008ApJ...683L..25P}, we corrected for the different IMF assumed in these papers compared to our work. }
\label{fig:DTD_all}
\end{figure}

\subsubsection{Delay-time distribution}
\label{sect:results:col:DTD}
We compute the delay-time distributions (DTDs) for the combined WD merger-induced SNe Ia channels described in \S~\ref{sect:results:col:chan}, and show the results in \F~\ref{fig:DTD_ind} and \F~\ref{fig:DTD_all} for the individual channels and the combined channels, respectively. The DTDs are expressed in units of SNuM, i.e., rate per century per $10^{10}\,\msun$ of solar mass. In \F\,\ref{fig:DTD_all}, the top and bottom panels correspond to flybys excluded and included, respectively. Thick blue (thin red) lines show the DTDs with WD kicks included (excluded). The time-integrated rates are given in the legends (errors are based on Poisson statistics). In \F\,\ref{fig:DTD_all}, the black solid lines with blue error regions show the observed DTD from \citet{2017ApJ...848...25M}, i.e., $\mathrm{DTD}\propto t^{-\alpha}$ with $\alpha=1.07^{+0.09}_{-0.08}$ and a Hubble time-integrated rate of $(1.3\pm0.1)\times 10^{-3} \, \msun^{-1}$. We also show, with dashed black lines with green error regions, the DTD of \citet{2008ApJ...683L..25P}. 

Comparing the individual channels (\F~\ref{fig:DTD_ind}), RLOF-induced mergers dominate the simulated rates at times $\gtrsim 1\,\mathrm{Gyr}$. The second contribution comes from dynamically unstable systems, although the assumed efficiency in this channel is uncertain (cf. \S~\ref{sect:results:col:chan}). The semisecular regime and secular collisions contribute only little to the simulated rates.

The DTDs from the combined channels in our simulations (\F~\ref{fig:DTD_all}) are relatively flat, i.e., flatter than $\propto t^{-1}$ as often found in binary population synthesis studies (e.g., \citealt{2012A&A...546A..70T}), and observations (e.g., \citealt{2012MNRAS.426.3282M}). Without WD kicks and flybys, the time-integrated rates are of the order of $10^{-5}\,\msun^{-1}$, somewhat higher compared to previous studies \citep{2013MNRAS.430.2262H,2018A&A...610A..22T}. The latter can be explained by the inclusion of RLOF-induced mergers in our work, which dominate the rates. Including WD kicks (with flybys still excluded) increases the rates by a factor of $\approx 1.8$ to $\approx 8\times 10^{-5}\,\msun^{-1}$. When flybys are included, the boost by WD kicks is a factor of $\approx 1.7$. All in all, the difference in rates between the `vanilla' case without flybys and WD kicks, and the `full-fledged' case with flybys and WD kicks, is a factor of $\approx 2.5$. 

For reference, studies of isolated binary evolution find rates which are a factor of $\sim 10$ lower compared to the observed rate of $\approx 10^{-3} \, \msun^{-1}$ (e.g., \citealt{2011MNRAS.417..408R,2014A&A...563A..83C}). For quadruples, \citet{2018MNRAS.478..620H} find a combined rate for 2+2 and 3+1 systems which is about $5\times 10^{-6} \, \msun^{-1}$ (taking into account flybys, but not WD kicks).

\section{The impact of WD kicks on more compact systems}
\label{sect:discussion:compact}
As described in \S~\ref{sect:meth:IC:orbits}, we restricted to triples with non-interacting binaries in the absence of a tertiary companion, i.e., triples with inner binaries initially satisfying $a_1(1-e_1^2)<12\,\au$. Here, we succinctly estimate the impact of WD kicks on tighter systems. A detailed investigation of the tighter triples is beyond the scope of this paper (see, e.g., \citealt{2019MNRAS.484.1506L} for a study of the implications of the effects of kicks alone, i.e., without secular/stellar evolution, in hierarchical triple systems, in particular in the context of shrinking the inner binary).

\begin{figure}
\center
\includegraphics[scale = 0.48, trim = 5mm 0mm 0mm 0mm]{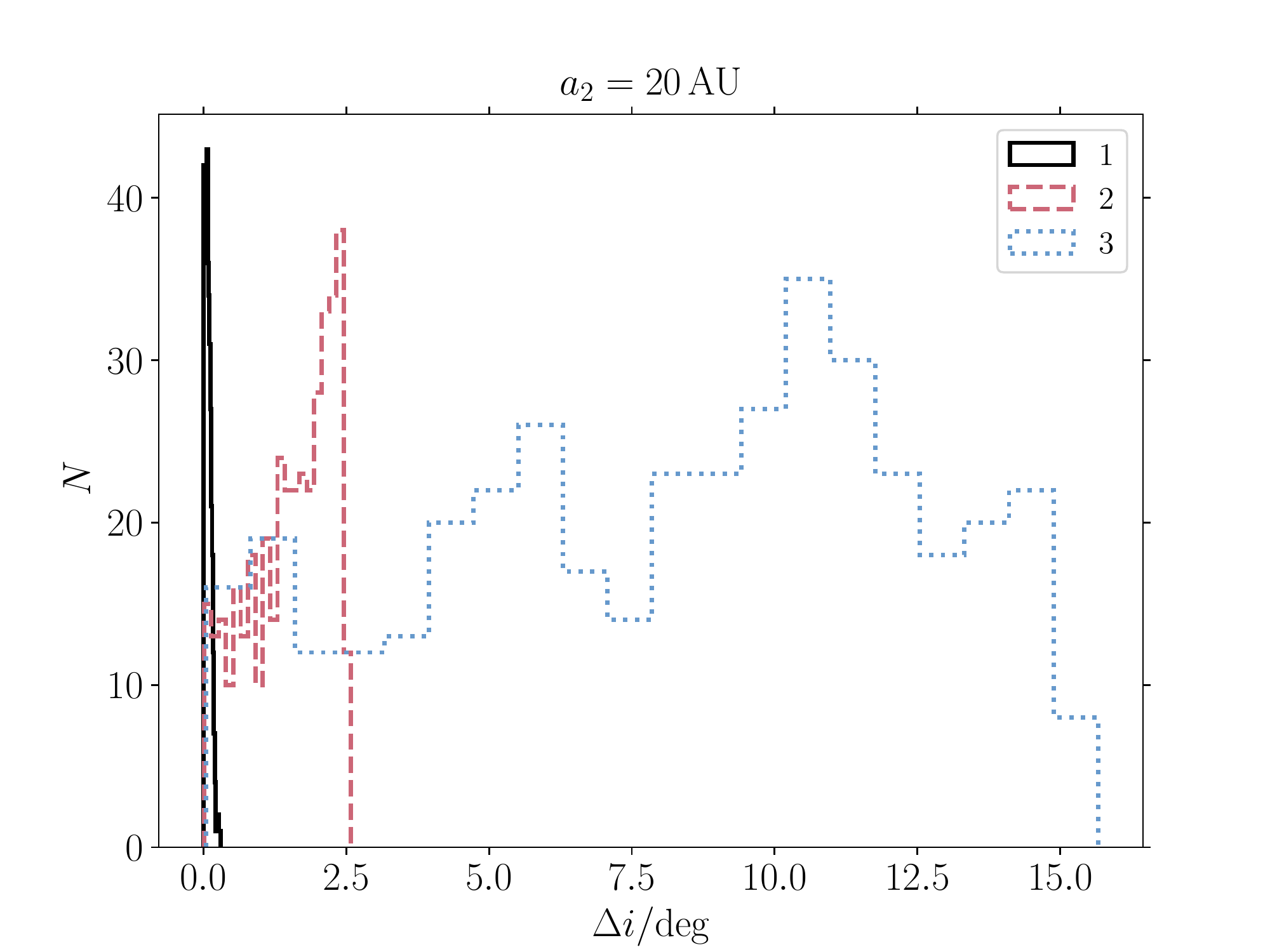}
\includegraphics[scale = 0.48, trim = 5mm 0mm 0mm 0mm]{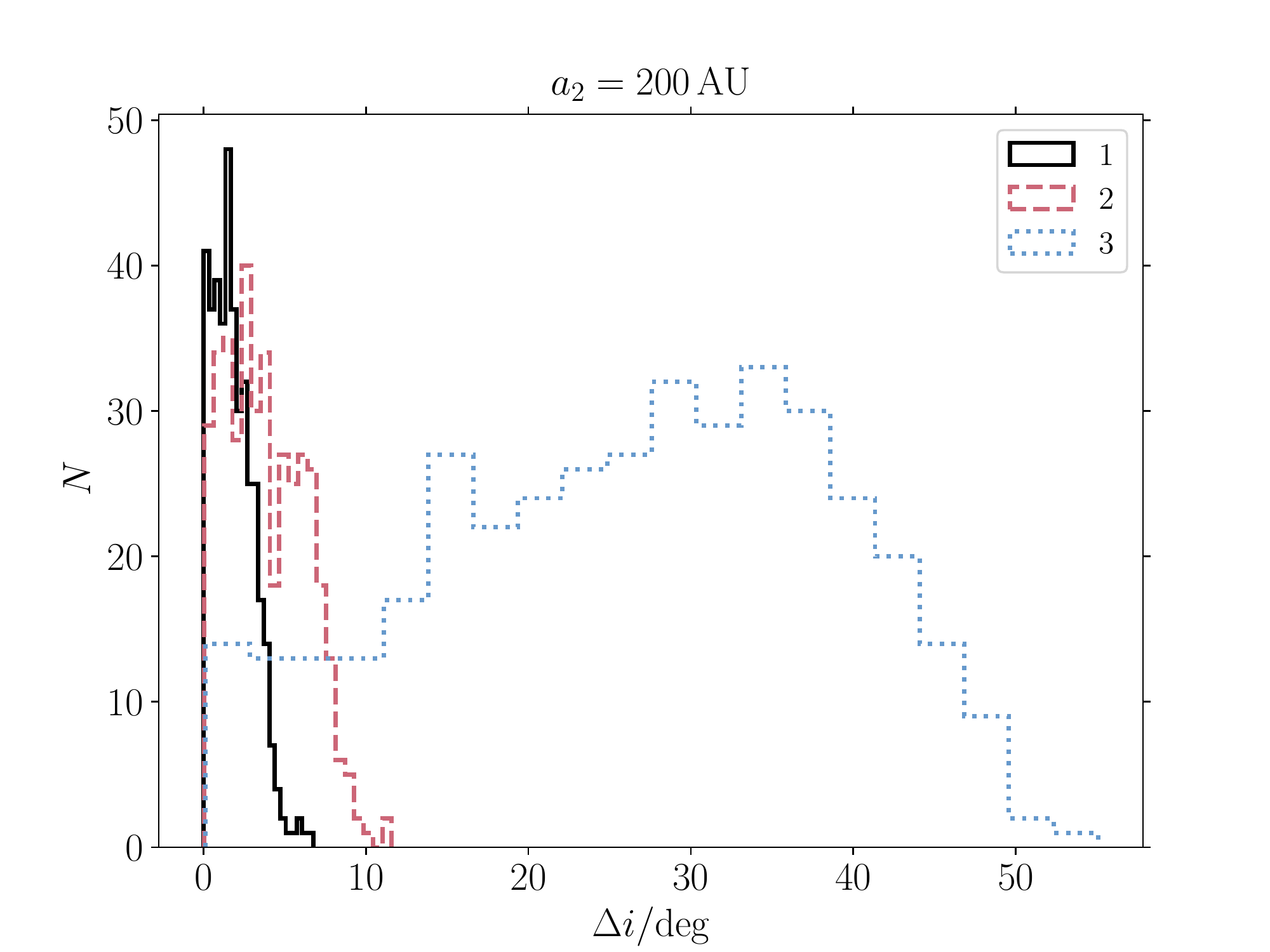}
\caption { Change in mutual inclination for a triple with initially $a_1=2\,\au$, $a_2=20\,\au$ (top panel) or $a_2=200\,\au$ (bottom panel), $e_1=0.1$, $e_2=0.3$, $\Delta i=0^\circ$, $\omega_1=\omega_2=0^\circ$. Initial masses are $m_1=0.6\,\msun$, $m_2=1.5\,\msun$, and $m_3=1\,\msun$. WD kicks are subsequently applied when the primary, secondary, and tertiary star becomes a WD, respectively. Shown is the distribution of the inclination change during each stage for a number of different realizations. Adiabatic mass loss is neglected.}
\label{fig:delta_i_tight}
\end{figure}

In \F~\ref{fig:delta_i_tight}, we show the change in mutual inclination for a triple with initially $a_1=2\,\au$, $a_2=20\,\au$ (top panel) or $a_2=200\,\au$ (bottom panel), $e_1=0.1$, $e_2=0.3$, $\Delta i=0^\circ$, $\omega_1=\omega_2=0^\circ$. The initial masses are $m_1=0.6\,\msun$, $m_2=1.5\,\msun$, and $m_3=1\,\msun$. We subsequently consider the effect of a WD kick on the system when the primary, secondary, and tertiary star becomes a WD, in each case a WD with a mass of $0.6\,\msun$. For simplicity, we neglect the effect of adiabatic mass loss on the orbits. For each stage, we sample multiple kick velocities using the same distribution as in \S~\ref{sect:meth:kick}, i.e., with $\sigma_\mathrm{k} = 0.5\,\mathrm{km\,s^{-1}}$ \citep{2018MNRAS.480.4884E}.

In the case of a compact triple ($a_2=20\,\au$; top panel of \F~\ref{fig:delta_i_tight}), $\Delta i$ after the first kick is very small, i.e., significantly less than $1^\circ$. The second kick can reach $\Delta i = 2.5^\circ$, with up to $\Delta i \approx 15^\circ$ when the tertiary receives a kick. These inclinations can be considered small. When the triple is less compact ($a_2=200\,\au$; bottom panel of \F~\ref{fig:delta_i_tight}), the inclination changes are significantly larger, up to $\approx 50^\circ$. Such a large change in the mutual inclination may well imply that a significant fraction of systems can become dynamically active at late times. However, for wider triples, it should also be taken into account that LK oscillations can be suppressed by short-range forces (e.g., \citealt{2015MNRAS.447..747L}). These compact systems should be the focus of additional studies (see \S~\ref{sect:conclusions}). 

\section{Conclusions}
\label{sect:conclusions}
Using detailed simulations, we investigated whether WD natal kicks and passing stars could enhance the rate of SNe Ia arising from CO WD mergers in hierarchical triple stars. Our main conclusions are as follows.

\medskip \noindent 1. We carried out four sets of simulations, taking each combination of excluding or including WD natal kicks, and flybys. Without WD kicks, the majority of systems ($\sim 0.8$) do not interact during the evolution. However, when WD kicks are included, only $\sim $ 0.1-0.2 of systems do not interact. Instead, a large fraction of systems becomes unbound ($\sim$ 0.5-0.6), or dynamically unstable ($\sim 0.2$). Flybys slightly increase the occurrence of RLOF, dynamical instability, and the onset of the semisecular regime in which the double-averaged approximation breaks down. 

\medskip \noindent 2. Although we focused on relatively wide triples in which the inner system does not interact in the absence of the tertiary star (inner orbit semilatus rectum $a_1[1-e_1^2] > 12\,\au$), in a significant fraction of systems ($\sim 0.1$), RLOF is triggered. Of the systems undergoing RLOF, the inner binary is highly eccentric for a substantial fraction  ($e> 0.9$ for $\sim 30\%$ of the RLOF systems). 

\medskip \noindent 3. We identified the following channels that can result in CO WD mergers, and, therefore, potentially SNe Ia: direct collisions due to extremely high eccentricities according to the doubly-averaged equations of motion, collisions following dynamical instability, collisions due to evolution in the semisecular regime, and circular mergers of WDs in tight orbits driven by gravitational wave emission, after binary interaction triggered by the tertiary (i.e., RLOF-induced mergers). The delay-time distributions (DTDs) of the combined channels are flatter than $\propto t^{-1}$. 

\medskip \noindent 4. Our time-integrated rate without WD kicks and flybys is $\approx 5 \times 10^{-5} \,\msun^{-1}$, somewhat higher compared to previous works due to the inclusion of RLOF-induced mergers. Considered separately, WD kicks and flybys increase the time-integrated rates by a factor of $\sim 2$ and $\sim 1.4$, respectively. Combined, the two effects give a boost of a factor of $\sim 2.5$, with a time-integrated rate of $\approx 1.1\times 10^{-4}\,\msun^{-1}$. Despite the significant boost from WD kicks and flybys, the predicted rates are still an order of magnitude below the observed rate of $\sim 10^{-3} \, \msun^{-1}$. 

In this paper, we addressed two new aspects in the modeling of WD mergers from triple systems: flybys and WD kicks. These effects increase the predicted WD merger rates, although not to the level of the observed SNe Ia rate. However, large systematic uncertainties still remain. Most critically, we did not consider the evolution of compact triples, i.e., triples with inner binary semilatus recti $a_1(1-e_1^2) < 12 \,\au$. We expect that there may be an up to order-of-magnitude impact in the predicted merger rates from these systems, which were omitted in this study (and in previous works). In addition, we assumed an optimistic but highly uncertain WD collision efficiency of 10\% for the systems that become dynamically unstable (see \S~\ref{sect:results:col:chan}). This efficiency should be determined accurately (e.g., using direct $N$-body integrations) in future study. Similar work should address the outcome of unbound orbits due to WD kicks. Another unaddressed problem in the theoretical modeling is the impact of WD kicks on WD mergers in higher-multiplicity systems, although it is unlikely that this effect would increase the merger rate by more than factors of a few. Furthermore, uncertainties also remain in the modeling of WD mergers from (isolated) binary systems (e.g., \citealt{2018arXiv180203125L}), and in the characteristics of the WD population in multiple systems.

\section*{Acknowledgements}
We thank the anonymous referee for helpful comments. A.S.H. gratefully acknowledges support from the Institute for Advanced Study, and the Martin A. and Helen Chooljian Membership. T.A.T. is supported in part by a Simons Foundation Fellowship, an IBM Einstein Fellowship from the Institute for Advanced Study, and by NSF grant 1313252.

\bibliographystyle{yahapj}

\bibliography{literature}


\end{document}